%% file: extended.tex
\documentclass[sigconf,nonacm]{acmart}

\input{preamble}
\input{vldb_macros}
\input{macros}

\togglefalse{highlighted}\toggletrue{extended}%

\begin{document}
\title{Reasoning about Transactional Isolation Levels with \textsc{Isolde}}

\author{Manuel Barros}
\affiliation{%
  \institution{Carnegie Mellon University}
  \city{Pittsburgh}
  \country{USA}
}
\email{mbarros@cmu.edu}

\author{Eunsuk Kang}
\affiliation{%
  \institution{Carnegie Mellon University}
  \city{Pittsburgh}
  \country{USA}
}
\email{eskang@cmu.edu}

\author{Alcino Cunha}
\affiliation{%
  \institution{INESCTEC \& U. Minho}
  \city{Braga}
  \country{Portugal}
}
\email{alcino@di.uminho.pt}

\author{José Pereira}
\affiliation{%
  \institution{INESCTEC \& U. Minho}
  \city{Braga}
  \country{Portugal}
}
\email{jop@di.uminho.pt}

\begin{abstract}
  Most databases can be configured to operate under \emph{isolation
  levels} weaker than \emph{serializability}. These enforce fewer
  restrictions on the concurrent access to data and consequently
  allow for more performant implementations. While formal frameworks for
  rigorously specifying isolation levels exist, reasoning about the
  semantic differences between specifications remains notoriously difficult.

  This paper proposes  a tool --- Isolde --- that can automatically
  generate examples that are allowed by an isolation level but
  disallowed by another. This simple primitive unlocks a range of
  useful reasoning tasks, including checking equivalence between
  definitions, and verifying (by refutation) subtle semantic
  properties of isolation levels.
  For example, Isolde allowed us to easily and automatically
  reproduce a famously elusive result from the literature and to
  discover a previously unknown bug in the alternative specification
  of a standard isolation level used in a state-of-the-art isolation checker.
\end{abstract}

\maketitle

\pagestyle{\vldbpagestyle}
\begingroup\small\noindent\raggedright\textbf{PVLDB Reference Format:}\\
\vldbauthors. \vldbtitle. PVLDB, \vldbvolume(\vldbissue): \vldbpages,
\vldbyear.\\
\href{https://doi.org/\vldbdoi}{doi:\vldbdoi}
\endgroup
\begingroup
\renewcommand\thefootnote{}\footnote{\noindent
  This work is licensed under the Creative Commons BY-NC-ND 4.0
  International License. Visit
  \url{https://creativecommons.org/licenses/by-nc-nd/4.0/} to view a
  copy of this license. For any use beyond those covered by this
  license, obtain permission by emailing
  \href{mailto:info@vldb.org}{info@vldb.org}. Copyright is held by
  the owner/author(s). Publication rights licensed to the VLDB Endowment. \\
  \raggedright Proceedings of the VLDB Endowment, Vol. \vldbvolume,
  No. \vldbissue\ %
  ISSN 2150-8097. \\
  \href{https://doi.org/\vldbdoi}{doi:\vldbdoi} \\
}\addtocounter{footnote}{-1}\endgroup

\ifdefempty{\vldbavailabilityurl}{}{
  \vspace{.3cm}
  \begingroup\small\noindent\raggedright\textbf{PVLDB Artifact Availability:}\\
  The source code, data, and/or other artifacts have been made
  available at \url{\vldbavailabilityurl}.
  \endgroup
}

\iftoggle{highlighted}{%
  \linenumbers
}

  \input{sections/introduction}  \input{sections/background}  \input{sections/showcase}  \input{sections/technique}%
  \input{sections/evaluation}  \input{sections/relatedwork}  \input{sections/conclusion}  \bibliographystyle{ACM-Reference-Format}
  \bibliography{references}%

\iftoggle{extended}{%
  \clearpage
  \appendix
  \section*{Appendix}
  \input{sections/appendix/appendix-algo}
  \input{sections/appendix/appendix-evaluation}
}

\end{document}

%% file: preamble.tex
\usepackage[group-minimum-digits=4,per-mode=fraction]{siunitx}
\usepackage{amsmath}
\usepackage{longtable}
\usepackage{csquotes}
\usepackage{amsfonts}
\usepackage{tikz}
\usepackage{algorithm}
\usepackage{algpseudocodex}
\usepackage{tikz-cd}            
\usepackage{subcaption}
\usepackage{centernot}
\usepackage{soul}
\usepackage[shortlabels]{enumitem}
\usepackage{booktabs}
\usepackage{multirow}
\usepackage{minted}
\usepackage{makecell}
\usepackage{enumitem}
\usepackage{etoolbox}
\usepackage{afterpage}
\usepackage{pdfpages}
\usepackage{lineno}
\usepackage{listings}

%% file: vldb_macros.tex
\newcommand\vldbdoi{XX.XX/XXX.XX}
\newcommand\vldbpages{XXX-XXX}
\newcommand\vldbvolume{14}
\newcommand\vldbissue{1}
\newcommand\vldbyear{2020}
\newcommand\vldbauthors{\authors}
\newcommand\vldbtitle{\shorttitle}
\newcommand\vldbavailabilityurl{URL_TO_YOUR_ARTIFACTS}
\newcommand\vldbpagestyle{plain}

%% file: macros.tex
\NewDocumentCommand{\mvis}{}{\mathit{vis}}
\NewDocumentCommand{\mar}{}{\mathit{ar}}
\NewDocumentCommand{\vis}{}{$\mvis{}$}
\NewDocumentCommand{\arb}{}{$\mar{}$}
\NewDocumentCommand{\mco}{}{\mathit{co}}
\NewDocumentCommand{\co}{}{$\mco{}$}

\newcommand\relatedBy[3]{#2 \xrightarrow{#1} #3}
\newcommand\spec[2]{\left\{ #1, \overline{#2} \right\}}

\newcounter{history}

\newcounter{transaction}[history]

\newenvironment{history}
{
  \refstepcounter{history}
  \newcommand{\txn}[1]{\stepcounter{transaction}T_{\thetransaction} &: ##1 \\}
  \renewcommand{\r}[2]{\text{r}(\mathtt{##1}, ##2)\ }
  \newcommand{\w}[2]{\text{w}(\mathtt{##1}, ##2)\ }
\[\aligned}{\endaligned\]}

\newtoggle{highlighted}
\newtoggle{extended}

\newcommand{\revised}[1]{%
  \iftoggle{highlighted}{{\color{blue}#1}}{#1}%
}

\newcommand{\revisedinline}[1]{%
  \iftoggle{highlighted}{\textcolor{blue}{#1}}{#1}%
}

\newcommand{\onlyclean}[1]{%
  \iftoggle{extended}{}{#1}%
}

\newcommand{\cleanorextended}[2]{
  \iftoggle{extended}{#2}{#1}%
}

\ifdefined\mode\else\def\mode{paper}\fi

\algnewcommand\algorithmicbreak{\textbf{break}}
\algnewcommand\Break{\State \algorithmicbreak}

%% file: sections/introduction.tex
\revised{
  \section{Introduction}\label{sec:introduction}

  Transactional techniques are key to handling concurrency and faults
  in database systems and applications~\cite{Helland2024-nn}. Systems
  often aim at \emph{serializability} --- that is, transactions
  behave as if executed atomically in some serial
  order~\cite{papadimitriouSerializabilityConcurrentDatabase1979} ---
  which greatly simplifies the tasks of developing applications and
  ensuring their correctness.

  Enforcing serializability does, however, impose concurrency control
  and distributed coordination
  overheads~\cite{Harizopoulos2018-qd,Zhou2025-cu}. Therefore, system
  developers have historically sought weaker \emph{isolation
  levels}~\cite{berensonCritiqueANSISQL1995}. More recently, weaker
  definitions have also been used to maximize availability and
  performance in distributed database
  systems\,\cite{DeCandia2007-ga,Lloyd2011-tq,Sovran2011-hv,Lloyd2013-di,Bailis2016-sp,Akkoorath2016-nb,Spirovska2019-ka,Liu2024-cc}.
  Weaker levels, however, allow for \emph{concurrency anomalies} that
  would not be possible in a sequential system, and are thus much
  harder for application developers to reason about, particularly if
  these levels are diverse and ambiguously defined or differ only in
  subtle details\,\cite{S_Elnikety_undated-gv,Ardekani2013-og}. In
  fact, Phil Bernstein, while summing up fifty years of research on
  transactions, has identified the need for \emph{principled
  approaches} to weaker isolation as a key outstanding challenge for
  database research~\cite{Bernstein2025-dj}.

  Formally defining isolation levels is a prevalent approach to
  addressing this complexity~\cite{adyaGeneralizedIsolationLevel2000,
    ceroneFrameworkTransactionalConsistency2015,
    crooksSeeingBelievingClientCentric2017,
    biswasComplexityCheckingTransactional2019,
    kakigowthamAloneTogetherCompositional2017,
  szekeresMakingConsistencyMore2018a}. Beyond offering developers a
  rigorous account of the consistency guarantees provided by each
  level, formal definitions serve as the theoretical foundation for
  tools that detect isolation-related bugs. They have been used to
  implement both static and dynamic analyses for applications running
  under weak isolation~\cite{biswasMonkeyDBEffectivelyTesting2021,
    ganIsoDiffDebuggingAnomalies2020,
    nagarAutomatedDetectionSerializability2018,
  bouajjaniDynamicPartialOrder2023}, and to construct checkers that
  verify whether databases adhere to their intended isolation
  levels~\cite{kingsburyElleInferringIsolation2020a,
    zhangViperFastSnapshot2023, liuPlumeEfficientComplete2024,
  biswasComplexityCheckingTransactional2019, Moeldrup2025}.

  However, relatively little attention has been paid to validating
  and reasoning about the isolation level specifications themselves.
  This gap matters for several communities: researchers who design
  novel isolation
  guarantees~\cite{bailisScalableAtomicVisibility2016}; those who
  re-specify existing levels~\cite{liuPlumeEfficientComplete2024,
    crooksSeeingBelievingClientCentric2017,
    ceroneFrameworkTransactionalConsistency2015,
  ceroneAnalysingSnapshotIsolation2018, liuROLANewDistributed2018};
  and those who prove semantic properties of known
  levels~\cite{feketeReadonlyTransactionAnomaly2004, Moeldrup2025,
    bouajjaniDynamicPartialOrder2023,
  ganIsoDiffDebuggingAnomalies2020}. The last concern is especially
  prevalent among developers of isolation-related verification tools,
  who frequently rely on custom properties of isolation definitions
  to justify algorithmic optimizations~\cite{Moeldrup2025,
    bouajjaniDynamicPartialOrder2023, ganIsoDiffDebuggingAnomalies2020,
  liuPlumeEfficientComplete2024}.

  Because of the inherently complex nature of concurrent systems,
  these are all non-trivial tasks that require careful thinking, and
  whose results need to be backed by formal
  proofs~\cite{liuPlumeEfficientComplete2024,
  crooksSeeingBelievingClientCentric2017, liu2019}. The difficulty is
  compounded when properties span multiple specification frameworks,
  as is the case when translating isolation definitions across formalisms.
  As a concrete example, consider the recent efforts by the authors
  of the Plume isolation checker~\cite{liuPlumeEfficientComplete2024}
  in designing novel specifications for standard isolation levels.
  The new definitions were necessary  to develop an efficient
  isolation checker that is complete (i.e., does not miss violations)
  and that produces understandable explanations of reported
  violations. Guaranteeing the correctness of these definitions
  required an iterative process of specification, attempted
  equivalence proof, and revision --- which was repeated until the
  proofs were successful.

  In this work, we present an automated approach for reasoning about
  isolation levels. Specifically, we address the problem of
  \emph{history synthesis}: given two isolation level specifications
  $A$ and $B$, the goal is to verify if there is any observable
  database behavior (henceforth, a \emph{history}), within a
  configurable bound that limits the size of histories, that is
  allowed by $A$ but disallowed by $B$ and, if there is, synthesize
  one such history.

  A lot of interesting properties can be verified by solving one or
  more such history synthesis problems. Notably, we can check if $A$
  implies $B$ ($A \Rightarrow B$), i.e., if the set of histories
  allowed by $A$ is a subset of the histories allowed by a $B$.
  Namely, finding a history allowed by $A$ but disallowed by $B$
  serves as a concrete proof (by refutation) that $A$ \emph{does not}
  imply $B$. Furthermore, one can check the equivalence between two
  definitions by verifying if both imply the other. This is useful,
  for instance, when designing an alternative specification for an
  existing isolation level.

  Besides serving as counterexamples to assertions comparing
  different isolation level definitions, synthesized histories  allow
  visualizing the concrete behavior that distinguishes those
  definitions, which helps building
  an intuition behind their semantical differences.

  Because the history synthesis problem is fundamentally expressed as
  a higher-order logic formula, it cannot be directly solved by traditional
  model-finding and model-checking approaches, such as
  Alloy~\cite{jacksonSoftwareAbstractionsLogic2012} or
  TLA+~\cite{lamport2002specifying}, which support only first-order
  logic. Furthermore, even under small bounds on history size, the
  set of possible histories is immense, making exhaustive enumeration
  techniques infeasible in practice. A practical synthesis technique
  must be efficient enough to support rapid experimentation, allowing
  researchers to test candidate specifications and obtain concrete
  feedback without the overhead of a full formal proof attempt. To
  this end, we propose a solution inspired by a program synthesis
  technique known as Counterexample-Guided Inductive Synthesis
  (CEGIS)~\cite{solar-lezamaCombinatorialSketchingFinite2006}, which
  we implemented in a tool called Isolde.

  Isolde is sound (i.e., synthesized histories are guaranteed to
  satisfy $A$ and violate $B$), and complete for the given bound
  (i.e., it is guaranteed to synthesize a history if some satisfying
  history exists within this bound). In practice, this means that
  successfully synthesizing a history allowed by $A$ and disallowed
  by $B$ serves as proof that $A \not\Rightarrow B$, but failing to
  do so does \emph{not} prove $A \Rightarrow B$ for arbitrarily sized
  histories (a counterexample could exist for a higher
  bound)\footnote{\revised{Throughout the paper, we reserve the word
      \emph{prove} for when Isolde is able to synthesize history, which
      constitutes
      a genuine refutation of an implication; we use terms like
      \emph{suggest} or \emph{support} to refer to properties which are
      supported by Isolde's UNSAT results, but not necessarily proven for
  arbitrarily sized histories, given the tool's bounded nature.}}.
  This bounded nature of
  Isolde, however, does not compromise its usefulness, since most
  isolation anomalies can be triggered with few transactions
  performing just a couple
  operations~\cite{cuiUnderstandingTransactionBugs2024}.

  The core technique behind Isolde's history synthesis is
  framework-agnostic: it supports isolation specifications given in
  most state-of-the-art axiomatic frameworks. In particular, a
  history synthesis problem involve specifications of different
  formal frameworks. This allows direct comparison of specifications
  across different frameworks.

  To showcase the usefulness of Isolde, we apply it to two problems
  taken from the literature. Firstly, we show how it could have been
  used to reduce and validate the specification effort of the authors
  of Plume in designing their alternative isolation level
  specifications. In fact, Isolde allowed us to uncover a
  specification mistake in one of their definitions. Secondly, we use
  it to reproduce a famously elusive theoretical result about the
  Snapshot Isolation level~\cite{feketeReadonlyTransactionAnomaly2004}.


  The rest of this paper is structured as follows. We start by
  providing some background on the formalization of isolation levels,
  using the framework
  by~\citet{biswasComplexityCheckingTransactional2019} as a concrete
  example of a formalism for defining isolation based on abstract
  executions (Section~\ref{sec:background}). Then, we introduce Isolde
  and showcase its usefulness by applying it to two problems taken from
  the literature (Section~\ref{sec:showcase})\footnote{The extended
  version of this paper includes a third case study of Isolde.}. We
  then describe in
  detail our synthesis technique (Section~\ref{sec:technique}). After
  that, we present the results of our evaluation of Isolde's
  performance (Section~\ref{sec:evaluation}). Lastly, we go over
  related work (Section~\ref{sec:related-work}) before some concluding
remarks (Section~\ref{sec:conclusion}).}

%% file: sections/background.tex
\section{Defining isolation}\label{sec:background}

Most formal frameworks for specifying isolation follow a paradigm
based on \emph{abstract executions}. For a concrete example of one of
these formalisms, we describe in detail the formal framework for
specifying isolation proposed
by~\citet{biswasComplexityCheckingTransactional2019}. At the end of
this section, we formalize the abstract execution paradigm, and
briefly go over how it applies to two other popular frameworks.

\subsection{Specifying isolation using a commit-order}

\citeauthor{biswasComplexityCheckingTransactional2019}~\cite{biswasComplexityCheckingTransactional2019}
model a transaction as a tuple $(O, po)$, where $O$ is a set of
\emph{operations} and $po$ (short for \enquote{program order}) is a
strict total order on $O$, representing the order of operations
issued by the transaction. The set of all operations is given by
\[
  \text{Op} = \{\ \text{read}_i(\mathtt{x}, v), \:
    \text{write}_i(\mathtt{x}, {v}) \, \mid \, i \in \text{OpId}, \:
  \mathtt{x} \in \text{Obj}, \: v \in \text{Val} \ \}
\]
We abbreviate operation names to \enquote{r} and \enquote{w} to
represent read and write operations, respectively. Obj and Val
correspond to the sets of all available database objects and values,
respectively. A special transaction $T_0$ is assumed to write the
initial value of all objects. Here, we assume that $\text{Val} =
\mathbb{Z}$ and that all objects are initially mapped to $0$. In a
given history, transactions all have disjoint sets of operations.
Semantically equivalent operations (e.g., writing the same value to
the same object) are distinguished by their operation identifiers taken
from OpId. When representing histories, we omit these identifiers.

A \emph{history} represents the observable result of a given database
execution. It is given by a tuple $(T, \mathit{so})$ in which $T$ is
a set of transactions assumed to have successfully committed, and
$\mathit{so}$ --- short for \emph{session order} --- is a partial
order over $T$ which effectively orders transactions issued
sequentially in the same client session. The initial transaction
$T_0$ is ordered before all other transactions in $\mathit{so}$.


A database \emph{execution} is a tuple $(H, \mathit{co})$, which
extends a history ($H$) with a total order $\mathit{co}$ (short for
\emph{commit order}) over $H$'s transactions, which represents the
order in which transactions commit. This is also sometimes referred
to as the \emph{arbitration} relation, since it reveals how the
database \enquote{chooses} to order concurrent transactions.
Naturally, $T_0$ precedes all other transactions in \co{}.

An isolation level is specified as a constraint on an execution. A
level specified by a constraint $M$ allows a history $H$ iff there
exists a total order $\mathit{co}$ such that $\left(H,
\mathit{co}\right)$ satisfies $M$, a property we denote as $H \models M$:
\[
  H \models M \iff \exists \mathit{co} \ldotp \left(H,
  \mathit{co}\right) \models M
\]

For conciseness, this formulation omits certain base semantic
constraints that every execution must satisfy, e.g., that
$\mathit{so} \subseteq \mathit{co}$ (meaning that the database must
respect the order of transactions issued by each client).

Additionally, when specifying constraints on executions,
\citeauthor{biswasComplexityCheckingTransactional2019} rely on a
relation $\mathit{wr} \subseteq T \times \text{Op}$, called
\emph{write-read}, which associates each read operation with the
transaction that wrote the value read. Inferring this relation from
the history alone is possible by assuming that each value is written
at most once. This is a common assumption across
frameworks~\cite{crooksSeeingBelievingClientCentric2017}. Any valid
commit order must respect the write-read relation: if a transaction
$T$ reads a value written by $S$ then $S$ must precede $T$ in \co{}.



\citeauthor{biswasComplexityCheckingTransactional2019} leverage this
framework to define several standard isolation levels. Each isolation
level essentially specifies a set of rules for when two transactions
must be related by $\mathit{co}$. As a concrete example, Read Atomic
requires that, if a transaction $t_3$ reads a value from object $x$
written by $t_1$, then every other transaction $t_2$ that writes to
$x$ and precedes $t_3$ in either $\mathit{wr}$ or $\mathit{so}$ must
be ordered \emph{before} $t_1$ in $\mathit{co}$. Formally:
\[
  \forall x, \: \forall t_1, t_2, t_3 \: \ldotp \: \left(
    \begin{array}{l c}
      t_1 \neq t_2                                           & \land \\
      \left(t_1, t_3\right) \in \mathit{wr}_x                & \land \\
      t_2 \in \text{writes}\left(x\right)                    & \land \\
      \left(t_2, t_3\right) \in \mathit{wr} \cup \mathit{so} &
    \end{array}
  \right) \implies
  \left(t_2, t_1\right) \in \mathit{co}
\]

The $\mathit{wr}_x$ relation maps each transaction $t$ that writes to
$x$ to all transactions $s$ that read the value written by $t$ to
$x$. The writes function produces the set of transactions that wrote
to a given object.

\revised{%
  \subsection{The Abstract Execution Paradigm}\label{sec:paradigm}

  The framework by \citet{biswasComplexityCheckingTransactional2019}
  follows the same general specification paradigm as most of the
  popular formal frameworks for specifying
  isolation~\cite{adyaGeneralizedIsolationLevel2000,
    ceroneFrameworkTransactionalConsistency2015,
  crooksSeeingBelievingClientCentric2017}. Namely, in all these
  frameworks, an isolation level is specified \emph{declaratively} as
  a first-order logic
  formula over abstract executions, characterizing which executions
  are valid under that level. An abstract execution is formed by two components:
  \begin{enumerate}
    \item \emph{A history.} The client-observable result of executing a set
      of transactions: the operations each transaction performed, the
      values they read and wrote, and the ordering constraints imposed
      by client sessions.

    \item \emph{A witness structure.} A set of auxiliary relations or
      structures, not directly observable by clients, that represent
      internal database decisions.
  \end{enumerate}
  For a history $H$ to be \emph{allowed} by a level $M$, there must
  exist a witness
  structure $W$ such that $(H, W)$ satisfies the constraint for $M$:
  \[
    H \models M \iff \exists W.\ (H, W) \models \phi_M
  \]
  where $\phi_M$ is the constraint characterising level $M$ under
  the given framework.

  Frameworks generally agree on the notion of a history, and differ
  primarily on the witness structure that is used as an abstraction
  for a database's internal decisions. In case of the framework
  by~\citeauthor{biswasComplexityCheckingTransactional2019}, the
  commit order is the witness structure. We now briefly show how two
  other frameworks follow this paradigm, which we henceforth refer to
  as the \emph{abstract execution} paradigm.

  \paragraph{\citet{ceroneFrameworkTransactionalConsistency2015}}
  \citeauthor{ceroneFrameworkTransactionalConsistency2015} use
  two witness relations: a \emph{visibility} relation
  (\vis), where $(t, s) \in \mvis$ means the writes of $t$ are
  observable by $s$; and an \emph{arbitration} relation (\arb),
  equivalent to the commit-order of
  \citeauthor{biswasComplexityCheckingTransactional2019}.
  Their
  formulation applies only to isolation levels to guarantee \emph{atomic
  visibility}: a transaction $T$ observes either all or none of
  the writes of any other transaction $S$.

  \paragraph{\citet{crooksSeeingBelievingClientCentric2017}}
  \citeauthor{crooksSeeingBelievingClientCentric2017} model
  executions as sequences of state-transitions triggered by
  transactions, where each database state maps objects to values.
  Each isolation level constrains the sequence of states that
  precedes each transaction. A history $H$ is allowed by a
  level $M$ if a state-transition sequence labeled by $H$'s
  transactions consistent
with $M$ exists.}

%% file: sections/showcase.tex
\section{A showcase of Isolde}\label{sec:showcase}

We have implemented our history synthesis technique in a tool called
Isolde\footnote{The source for Isolde is available at
\url{https://github.com/manebarros/isolde}.}. In this section we
showcase our tool's usefulness by applying it to
\cleanorextended{two}{three} different
problems taken from the literature.\onlyclean{\revisedinline{The
    extended version of this
work~\cite{extended_paper} goes over a third case study.}}

Before showing Isolde in action, it is useful to formalize the
problem it solves and introduce some notation. In reality, Isolde is
not restricted to comparing two isolation definitions. Instead, it
solves a more general problem: given two sets of isolation levels
$\left\{M_0, \ldots, M_m\right\}$ and $\left\{N_0, \ldots,
N_n\right\}$, and a finite universe of histories $\mathcal{S}$ (which
  is specified indirectly through a \emph{scope} that bounds some
history parameters, such as the number of transactions), verify if
there exists some history $H \in \mathcal{S}$ which is allowed by all
levels in $\left\{M_0, \ldots, M_m\right\}$ and disallowed by all
levels in $\left\{N_0, \ldots, N_n\right\}$, and, if so, produce one
such witness history. Formally:
\[
  \exists H \in \mathcal{S} \: \ldotp \: H \models M_0 \wedge \ldots
  \wedge H \models M_m \; \wedge \; H \not\models N_0 \wedge \ldots
  \wedge H \not\models N_n
\]

Henceforth, the following notation will be used to refer to such
synthesis problem:
\[
  \exists H \in \mathcal{S} \ldotp H \models \left\{ M_0, \ldots,
  M_m, \overline{N_0}, \ldots, \overline{N_n} \right\}
\]

Isolde is implemented as a Java library and currently accepts
specifications under the formalisms of
\citet{ceroneFrameworkTransactionalConsistency2015} and
\citet{biswasComplexityCheckingTransactional2019}. A synthesis
problem can contain definitions expressed in different formal
frameworks, allowing the comparison of specifications across formalisms.

Throughout the rest of this section, we show how Isolde can be used
to solve two problems of different nature. Firstly, we leverage it to
verify the equivalence of different specifications of isolation
levels. In particular, we verified the equivalence of the alternative
definitions of standard isolation levels proposed by
\citet{liuPlumeEfficientComplete2024} for the Plume isolation
checker. In this case,  Isolde allowed us to  a mistake in one of
their specifications. Secondly, we showcase how our synthesis
technique can be used to reason about subtle behaviors of isolation
levels. In particular, we used Isolde to automatically verify a known
non-trivial property about Snapshot Isolation involving read-only
transactions~\cite{feketeReadonlyTransactionAnomaly2004}.

\subsection{Verifying the equivalence of specifications}
\label{sec:equivalenceofspecs}
As was briefly mentioned in Section~\ref{sec:introduction}, history
synthesis allows us to test if one level $A$ implies another level
$B$. Namely, finding a history $H \models \left\{A,
\overline{B}\right\}$ is proof that $A$ \emph{does not} imply $B$.
While failing to synthesize such a history does \emph{not} prove the
implication (since Isolde's synthesis is necessarily bounded under a
finite scope), it constitutes strong supporting evidence for it
(assuming a relatively large scope).

This, in turn, can be used to assess how any two levels relate.
Particularly, given two isolation level specifications $A$ and $B$,
one of three outcomes holds:
\begin{enumerate}
  \item They both allow the same set of histories, in which case we
    say they are \emph{equivalent}:
    \[
      A \iff B
    \]
  \item They both allow (at least) a history that is disallowed by
    the other, in which case we say they are \emph{incomparable}:
    \[
      A \centernot\implies B \; \land \; B \centernot\implies A
    \]
  \item One of them allows for a proper subset of histories of the
    other, in which case we say that the former is \emph{stronger}
    than the latter. For instance, $A$ is stronger than $B$ iff the
    following holds:
    \[
      A \implies B \; \land \; B \centernot\implies A
    \]
\end{enumerate}

To infer how two specifications $A$ and $B$ relate, we can solve both
problems $\left\{A, \overline{B}\right\}$ and $\left\{B,
\overline{A}\right\}$. Failing to find either history suggests the
two specifications are equivalent; finding both histories is proof
that the two specifications are incomparable; and finding only one of
them suggests that one of the levels is stronger than the other.

\subsubsection{Verifying Plume's alternative isolation
specifications} As a concrete use case, consider the problem of
verifying if two isolation specifications are equivalent (i.e.,
correspond to the same isolation level). This was required by
\citet{liuPlumeEfficientComplete2024} when designing custom
specifications of four standard isolation levels: Cut Isolation
(\textsc{CI}), Read Committed (\textsc{RC}), Read Atomic
(\textsc{RA}), and Causal Consistency (\textsc{CC}). Their goal was
to completely characterize each level by a set of isolation anomalies
disallowed at that level. Their anomalies are essentially patterns of
database execution that reveal particular errors in the underlying
isolation-enforcing concurrency control mechanism. By characterizing
isolation levels based on fined-grained anomalies, and having their
checker look for occurrences of these patterns, they look to produce
understandable reports on the cause of each isolation violation. To
ensure that the new definitions were correct, the authors had to
prove their equivalence to existing definitions of the same levels.

Their definitions are given in the formal framework by
\citet{biswasComplexityCheckingTransactional2019}. They define a
total of fourteen \emph{Transactional Anomalous Patterns} (or TAPs),
named $\text{TAP}_a, \, \text{TAP}_b, \, \ldots, \, \text{TAP}_n$,
each specified as a formula on an execution $(H, \mco{})$. Each
isolation level is specified as a unique subset of these anomalies in
the following way. A history $H$ is allowed by a level $M$
characterized by the set of anomalies $\left\{A_0, \ldots,
A_x\right\}$ iff there is a commit-order \co{} such that $(H,
\mco{})$ does not contain any of $\left\{A_0, \ldots, A_x\right\}$:
\[
  H \models M \iff \exists \mco \ldotp (H, \mco) \not\models A_0
  \land \ldots \land (H, \mco) \not\models A_x
\]

For the new specifications to be correct (i.e., equivalent to
existing definitions), the set of anomalies that characterizes each
level must be \emph{complete} --- a history that can form an
execution that contains no anomaly is guaranteed to be allowed by the
level --- and \emph{sound} --- for every history allowed by the level
there is at least one anomaly-free execution. Arriving at such
definitions was reportedly challenging. The process for doing so
consisted in repeatedly designing the new definitions and attempting
to prove their equivalence to the formulations given by
\citet{biswasComplexityCheckingTransactional2019} until managing to
conclude the equivalence proofs.

\revised{%
  As is common with lightweight formal verification techniques, Isolde
  provides a stepping stone between informal verification and
  full-blown formal proof, by enabling automated formal guarantees in a
  limited universe of discourse. Isolde could have helped speed up the
  process of designing Plume's isolation specifications: Given a target
  isolation level $A$, and an attempt at an anomaly-based equivalent
  definition $B$, one could use Isolde to verify their equivalence as
  described above, and attempt to manually prove it only after
  successfully verifying it with Isolde under a significant scope.
}

We have used Isolde to verify the correctness of Plume's definitions
by comparing them with the definitions of
\citet{biswasComplexityCheckingTransactional2019}. In particular, we
focused on the alternative definitions of \textsc{RA} and
\textsc{CC}. To do so, for each of these levels, we verified the
equivalence between its axiomatic definition $M$ and its
anomaly-based counterpart $\textsc{Tap}M$ by following the method
described above, i.e., solving both synthesis problems $\left\{M,
\overline{\textsc{TapM}}\right\}$ and $\left\{\textsc{Tap}M,
\overline{M}\right\}$. We executed each problem with a maximum scope
of eight transactions, eight objects, and eight values.

In the case of \textsc{CC}, Isolde found no history for either
synthesis problem, thus supporting the equivalence between the two
definitions. For Read Atomic, however, Isolde is able to produce a
history allowed by $\textsc{TapRA}$ and disallowed by the axiomatic
definition RA (with synthesis problem $\left\{\textsc{TapRA},
\overline{\textsc{RA}}\right\}$), but not one allowed by $RA$ and
disallowed by $\textsc{Tap}RA$ (with synthesis problem $\left\{RA,
\overline{\textsc{Tap}RA}\right\}$). The history synthesized  by
Isolde for problem $\left\{\textsc{Tap}RA, \overline{RA}\right\}$
proves that the definitions are not equivalent. Namely, it proves
that $\textsc{TapRA}$ does not imply $\textsc{RA}$. The fact that
Isolde is not able to produce a history satisfying
$\left\{\textsc{RA}, \overline{\textsc{TapRA}}\right\}$ suggests that
the custom definition of Read Atomic proposed by
\citeauthor{liuPlumeEfficientComplete2024} is in fact weaker than the
definition given by \citeauthor{biswasComplexityCheckingTransactional2019}.

\paragraph{Understanding the difference between the two definitions}
Besides proving that the RA definitions are not equivalent,
, the synthesized histories give us insight into the differences between the
definitions, and provide guidance for fixing the anomaly-based specification.

By re-running the synthesis under
progressively smaller scopes, we obtain the following minimal
counterexample, with two transactions, one object, and two values:
\[
  T_1 : \, \text{r}(\mathtt{x}, 0)\ \text{w}(\mathtt{x}, 1) \quad
  \xrightarrow{\; \textit{so}\;} \quad T_2 : \, \text{r}(\mathtt{x}, 0)
\]

This history violates \textsc{RA} because $T_2$ fails to observe the
update by $T_1$, which precedes it in session order --- a direct
consequence of RA's axiomatic definition (showed in
Section~\ref{sec:background}) is that every transaction must observe
the writes of all transactions preceding it in \textit{wr} or \textit{so}.
Yet, the history is allowed by \textsc{TapRA}, since it admits a
commit-order which precludes all twelve anomalies
($\text{TAP}_a$ through $\text{TAP}_l$) that were conjectured to be
equivalent to \textsc{RA}. The problem is that all the anomalies
regarding transaction ordering involve at least two database
objects, and thus cannot capture this single-object
scenario\footnote{This problem has been confirmed to us by the
authors of Plume.}. A straightforward fix is
to add a catch-all anomaly consisting of the negation of the axiomatic
formula, mirroring the strategy the authors used for Causal Consistency.

\subsubsection{Verifying definition equivalence across frameworks}
Another scenario which requires reasoning about the equivalence
between isolation level definitions is when translating isolation
level definitions across formal frameworks. There is a plethora of
formalisms for characterizing isolation, each offering a unique
understanding of isolation levels. As described earlier, most of
these follow the same abstract execution paradigm for characterizing
isolation, which allows us to use our CEGIS algorithm to compare the
guarantees of definitions given in different formalisms.

In this regard, we have used Isolde to verify the equivalence between
the isolation level definitions proposed by
\citet{biswasComplexityCheckingTransactional2019}  with those by
\citet{ceroneFrameworkTransactionalConsistency2015}. Namely, we
tested the equivalence of definitions for five isolation levels: Read
Atomic, Causal Consistency, Prefix Consistency, Snapshot Isolation
and Serializable. We verified the equivalences using a maximum scope
of five transactions, five objects, and five values.


\subsection{Reasoning about isolation level
properties}\label{ssec:fekete_anomaly}
Besides comparing isolation levels, Isolde can be used to
automatically verify subtle properties about isolation levels.
\revised{%
  This is particularly useful for researchers that develop tools
  whose correctness depends on some custom properties on isolation levels.
  \cleanorextended{%
    To showcase this type of application, here we use Isolde to
    automatically reproduce a deceptive result about Snapshot
    Isolation first identified by
    \citet{feketeReadonlyTransactionAnomaly2004}. In the extended
    version of this paper~\cite{extended_paper} we additionally
    showcase a novel result about the Update Atomic isolation level
    proposed by \citet{ceroneFrameworkTransactionalConsistency2015}.
  }{%
    Here, we go over two problems. First, we use Isolde to verify a
    deceptive property about Snapshot Isolation first identified by
    \citet{feketeReadonlyTransactionAnomaly2004}. Then, we reason
    about the guarantees of the isolation level proposed by
    \citet{ceroneFrameworkTransactionalConsistency2015}.

    \subsubsection{A read anomaly under snapshot isolation}
}}

Consider the following property about
Snapshot Isolation:
\begin{quote}
  A history $H$ allowed by SI is serializable if the sub-history
  formed by $H$'s update transactions (transactions which perform at
  least a write operation) is serializable.
\end{quote}
For a long time, this was widely assumed to be
true~\cite{feketeReadonlyTransactionAnomaly2004}. The intuition was
that, under \textsc{SI}, read-only transactions are forced to read
from snapshots which contain only committed values. This seems to
imply that, if an SI history formed by update transactions is
serializable (i.e., it can be justified by some sequential execution
of those transactions) extending it with read-only transactions
(while still satisfying SI) will necessarily result in a set of
transactions which is also serializable.

However, this is, in fact, incorrect. As shown by
\citeauthor{feketeReadonlyTransactionAnomaly2004}~\cite{feketeReadonlyTransactionAnomaly2004},
it is possible for an \textsc{SI} database to produce a
non-serializable history whose sub-history containing all update
transactions is serializable. The reason for this is that the order
in which update transactions are serializable might not correspond to
the order in which those transactions actually commit. This can
result in read-only transactions consulting a database state which
would be impossible to achieve when considering a serial execution of
the update transactions.

We can use Isolde to automatically come to the same conclusion by
asking it to synthesize a history which refutes the property above.
That is, a history $H$ such that:
\begin{enumerate}
  \item $H$ is allowed by \textsc{SI};
  \item The sub-history formed by the update transactions of $H$ is
    serializable;
  \item $H$ is \emph{not} serializable.
\end{enumerate}

The conjunction of these three constraints can be translated to the
following Isolde specification:
\[
  \exists H \in \mathcal{S} \ldotp H \models \left\{ \textsc{SI},
  \textsc{UpdateSer}, \overline{\textsc{Ser}} \right\}
\]
Here, \textsc{SI} and \textsc{Ser} are the standard specifications of
Snapshot Isolation and Serializability. \textsc{UpdateSer} is an
additional assumption that enforces the constraints of \textsc{Ser}
only on the subset of transactions that perform at least one write operation.

The definition of \textsc{Ser} by
\citeauthor{biswasComplexityCheckingTransactional2019} requires a
\co{} edge connecting transactions $t_2$ and $t_1$ that write to an
item $x$ if there is a third transaction $t_3$ that reads the value
written to $x$ by $t_1$ and is ordered after $t_2$ in \co{}:
\[
  \forall x, \: \forall t_1, t_2, t_3 \: \ldotp \: \left(
    \begin{array}{l c}
      t_1 \neq t_2                            & \land \\
      \left(t_1, t_3\right) \in \mathit{wr}_x & \land \\
      t_2 \in \text{writes}\left(x\right)     & \land \\
      \left(t_2, t_3\right) \in \mathit{co}   &
    \end{array}
  \right) \implies
  \left(t_2, t_1\right) \in \mathit{co}
\]

\textsc{UpdateSer} enforces the same constraints as \textsc{Ser} only
on the transactions that perform at least a write operation. Thus, it
can be defined by extending the left side of the implication in the
formula above with the extra condition that $t_3$ performs a write operation.

For a concrete example of using our tool, we show we can actually
define this custom isolation criteria, use its definition for
specifying a synthesis problem, and finally use Isolde to solve that problem.

Users interact with Isolde through its Java API. Custom formulas and
expressions can be constructed using
Kodkod~\cite{torlakKodkodRelationalModel2007}, a popular Java library
that offers (1) an expressive specification language based on
first-order logic with relational-algebra constructs, and (2) a
model-finder for solving constraint-satisfaction problems specified
in this language. Listing~\ref{listing:updateSer} shows how we can
define \textsc{UpdateSer} to be used in a synthesis problem.

\iftoggle{extended}{%
  \input{sections/code/lstlisting/example}
}{%
  \input{sections/code/minted/example}
}

An isolation specification in the framework of
\citeauthor{biswasComplexityCheckingTransactional2019} is encoded as
function that takes in an abstract execution composed by a history
and a commit order, and produces a Kodkod formula enforces some
constraint on that abstract execution. As can be seen in this
snippet, the translation from formal notation to an actual encoding
in Isolde is quite straightforward. For the sake of space, we omit
the definition of the helper methods \texttt{writeTo}, which returns
the set of transactions in a history that write to a given object, and
\texttt{writeTxn}, which returns the set of update transactions in a
history. It is worth highlighting the meaning of the \texttt{join}
method, which performs relational join and is a powerful and
versatile operator when constructing formulas and expressions in
Kodkod. The remaining methods should be self-explanatory and are
omitted for clarity.

Listing~\ref{listing:synthesis} shows how one would use Isolde's API
to solve the desired problem \(\left\{ \textsc{SI},
\textsc{UpdateSer}, \overline{\textsc{Ser}} \right\}\). We use our
definition of \textsc{UpdateSer} given in
Listing~\ref{listing:updateSer} and predefined encodings of both
\textsc{SI} and \textsc{Ser}, which come packaged with Isolde in the
\texttt{Definitions} class. The parameters used to initialize the
scope correspond, respectively, to the number of transactions,
objects, and values.

\iftoggle{extended}{%
  \input{sections/code/lstlisting/using_definition}
}{%
  \input{sections/code/minted/using_definition}
}

\iftoggle{extended}{%
  \input{sections/code/lstlisting/output}
}{%
  \input{sections/code/minted/output}
}

Isolde successfully synthesis a history for this specification, which
it displays textually as is shown in Listing~\ref{listing:output}. By
default, Isolde uses integers as object identifiers. Mapping 0 to object
$y$ and 1 to $x$, the displayed history is equivalent to the following:

\begin{history}
  \txn{\r{y}{0} \w{y}{1}}
  \txn{\r{x}{0} \r{y}{1}}
  \txn{\r{x}{0} \r{y}{0} \w{x}{1}}
\end{history}

This history witnesses the kind of anomaly described by
\citeauthor{feketeReadonlyTransactionAnomaly2004}~\cite{feketeReadonlyTransactionAnomaly2004}.
Although the sub-history containing all the update transactions
(\(T_1, T_3\)) is serializable (by ordering \(T_1\) after \(T_3\)),
the full history is not: given \(T_2\)'s read on \(\mathtt{y}\), it
must be serialized \emph{after} \(T_1\); additionally, given
\(T_3\)'s read of the initial value on $\mathtt{y}$, it must be
serialized \emph{before} \(T_1\); however, \(T_2\)'s read on
\(\mathtt{x}\) invalidates this sequential execution.

Since the constraints $H \models \textsc{UpdateSer}$ and $H \models
\textsc{SI}$ both give rise to independent existential quantifiers on
binary commit-order relations, Isolde produces \emph{two} distinct
values of \co{}: one which justifies $H$ under \textsc{UpdateSer},
and a second one which justifies $H$ under \textsc{SI}. The orders in
which they are displayed are based on the orders in which they were
provided to Isolde during the construction of the spec. The values of
these relations help to understand this anomaly:
\begin{center}
  \begin{minipage}[t]{.35\linewidth}
    \textsc{UpdateSer}
    \vspace{.2cm}

    \begin{tikzpicture}
      \node[anchor=base] (t2) at (0, 0) {$T_3$};
      \node[anchor=base] (t1) at (0.5\textwidth, 0) {$T_1$};
      \node[anchor=base] (t3) at (\textwidth, 0) {$T_2$};
      \draw[arrows=->, dotted] (t2) to[out=25, in=-205] node[above]
      {\scriptsize{}\co{}} (t3);
      \draw[arrows=->, dotted] (t2) -- (t1);
      \draw[arrows=->, dotted] (t1) -- (t3);
    \end{tikzpicture}
  \end{minipage}
  \hspace{1cm}
  \begin{minipage}[t]{.35\linewidth}
    \textsc{SI}
    \vspace{.2cm}

    \begin{tikzpicture}
      \node[anchor=base] (t1) at (0, 0) {$T_1$};
      \node[anchor=base] (t3) at (0.5\textwidth, 0) {$T_3$};
      \node[anchor=base] (t2) at (\textwidth, 0) {$T_2$};
      \draw[arrows=->, dotted] (t1) to[out=25, in=-205] node[above]
      {\scriptsize{}\co{}} (t2);
      \draw[arrows=->, dotted] (t1) -- (t3);
      \draw[arrows=->, dotted] (t3) -- (t2);
    \end{tikzpicture}
  \end{minipage}
\end{center}

Note how the order in which \(T_1\) and \(T_3\) are serializable (on
the right-hand side execution) is different from the order in which
they commit in the \textsc{SI} execution.

\revised{%
  \paragraph{A note on choosing a scope}
  Much like other bounded solving tools, Isolde requires users to
  choose a scope when solving a
  synthesis problem. A scope has three dimensions: the number of
  transactions, objects, and
  values. Because isolation anomalies follow the small-scope
  hypothesis~\cite{cuiUnderstandingTransactionBugs2024},
  most satisfiable synthesis problems will be solved by a history
  with no more than five transactions
  (we elaborate on this in Section~\ref{sec:evaluation}).

  When using Isolde, we suggest starting with a relatively small scope
  (e.g., 3 transactions, 3 objects, and 3 values) and iteratively
  increasing it if Isolde returns UNSAT. As we
  show in Section~\ref{sec:evaluation}, for UNSAT problems, Isolde
  scales comfortably to five transactions and begins to
  show a significant performance slowdown only beyond six
  transactions. If Isolde does find a history, it is worth re-running the
  same problem under progressively
  smaller scopes to obtain more concise counterexamples. We
  suggest reducing each dimension in
  turn until any further reduction causes Isolde to return UNSAT,
  yielding a locally minimal witness.
  This is the strategy we followed when selecting the histories
  presented in this section.
}

\iftoggle{extended}{%
  \input{sections/extra/reasoning}
}{}

%% file: sections/code/lstlisting/example.tex
\begin{lstlisting}[float=tb, language=java, caption={Using Isolde's API to define \textsc{UpdateSer}.}, label=listing:updateSer]
Formula updateSer(BiswasExecution e) {
  Variable t1 = Variable.unary("t1");
  Variable t2 = Variable.unary("t2");
  Variable t3 = Variable.unary("t3");
  Variable x = Variable.unary("x");

  return Formula.and(
          t1.eq(t2).not(),
          e.history().wr(t1, x, t3),
          t2.product(t3).in(e.co()),
          t2.in(e.history().writeTo(x)),
          t3.in(e.history().writeTxn()))
      .implies(t1.in(t2.join(e.co())))
      .forAll(
          x.oneOf(e.history().objs())
              .and(t1.oneOf(e.history().txn()))
              .and(t2.oneOf(e.history().txn()))
              .and(t3.oneOf(e.history().txn())));
}
\end{lstlisting}

%% file: sections/code/minted/example.tex
\begin{listing}[tb]
  \begin{minted}{java}
Formula updateSer(BiswasExecution e) {
  Variable t1 = Variable.unary("t1");
  Variable t2 = Variable.unary("t2");
  Variable t3 = Variable.unary("t3");
  Variable x = Variable.unary("x");

  return Formula.and(
          t1.eq(t2).not(),
          e.history().wr(t1, x, t3),
          t2.product(t3).in(e.co()),
          t2.in(e.history().writeTo(x)),
          t3.in(e.history().writeTxn()))
      .implies(t1.in(t2.join(e.co())))
      .forAll(
          x.oneOf(e.history().objs())
              .and(t1.oneOf(e.history().txn()))
              .and(t2.oneOf(e.history().txn()))
              .and(t3.oneOf(e.history().txn())));
}
\end{minted}
  \caption{Using Isolde's API to define \textsc{UpdateSer}.}
  \label{listing:updateSer}
\end{listing}

%% file: sections/code/lstlisting/using_definition.tex
\begin{lstlisting}[float=tb, language=java, caption={Solving \(\left\{ \textsc{SI}, \textsc{UpdateSer},
  \overline{\textsc{Ser}} \right\}\).}, label=listing:synthesis]
Spec spec =
    biswas(Main::updateSer)
        .and(biswas(BiswasDefinitions::SI))
        .andNot(biswas(BiswasDefinitions::Ser))
        .build();

Scope scope = new Scope(3, 2, 2);

Synthesizer synth =
    new Synthesizer.Builder()
        .solver(SATFactory.MiniSat)
        .build();

SynthesisSolution sol = synth.synthesize(scope, spec);
System.out.println(synthesisResult);
\end{lstlisting}

%% file: sections/code/minted/using_definition.tex
\begin{listing}[tb]
  \begin{minted}{java}
Spec spec =
    biswas(Main::updateSer)
        .and(biswas(BiswasDefinitions::SI))
        .andNot(biswas(BiswasDefinitions::Ser))
        .build();

Scope scope = new Scope(3, 2, 2);

Synthesizer synth =
    new Synthesizer.Builder()
        .solver(SATFactory.MiniSat)
        .build();

SynthesisSolution sol = synth.synthesize(scope, spec);
System.out.println(synthesisResult);
\end{minted}
  \caption{Solving \(\left\{ \textsc{SI}, \textsc{UpdateSer},
  \overline{\textsc{Ser}} \right\}\).}
  \label{listing:synthesis}
\end{listing}

%% file: sections/code/lstlisting/output.tex
\begin{lstlisting}[float=tb, language=java, caption={Isolde's display of the history it synthesizes for
    \(\left\{ \textsc{SI}, \textsc{UpdateSer}, \overline{\textsc{Ser}}
  \right\}\).}, label=listing:output]
1: r(0,0) w(0,1)

2: r(0,1) r(1,0)

3: r(0,0) r(1,0) w(1,1)

Execution #1:
Commit Order:
3 -> 1 -> 2

Execution #2:
Commit Order:
1 -> 3 -> 2
\end{lstlisting}

%% file: sections/code/minted/output.tex
\begin{listing}[tb]
  \begin{minted}{text}
1: r(0,0) w(0,1)

2: r(0,1) r(1,0)

3: r(0,0) r(1,0) w(1,1)

Execution #1:
Commit Order:
3 -> 1 -> 2

Execution #2:
Commit Order:
1 -> 3 -> 2
    \end{minted}
  \caption{Isolde's display of the history it synthesizes for
    \(\left\{ \textsc{SI}, \textsc{UpdateSer}, \overline{\textsc{Ser}}
  \right\}\).}
  \label{listing:output}
\end{listing}

%% file: sections/extra/reasoning.tex
\subsubsection{Reasoning about Update Atomic}

In their paper describing an axiomatic framework for specifying isolation criteria, \citet{ceroneFrameworkTransactionalConsistency2015} propose a new isolation level --- called Update Atomic (\textsc{UA}) --- whose consistency guarantees sit between Read Atomic (\textsc{RA}) and Parallel Snapshot Isolation (\textsc{PSI}).
Namely, Update Atomic extends RA with a guarantee called \enquote{No Conflict}, which ensures that any two transactions that update an object in common are related by the visibility relation.
The motivation behind \textsc{UA} is to strengthn the guarantees of RA by preventing lost updates, but without imposing the causality constraints of PSI, which incur performance overheads in actual implementations.

We can use Isolde to verify that \textsc{UA} does prevent lost updates. To do this, we start by defining a lost update as a constraint on the abstract executions of \citet{ceroneFrameworkTransactionalConsistency2015}:
\begin{gather*}
	\textsc{LU} \doteq
	\exists t \in H \ldotp \ \exists x \in \text{Obj} \ldotp \  \mathit{writes}(t, x) \; \land \\
	\mathit{max}_{\mar{}}\left(\mar{}^{-1}(t) \cap \{s \mid \mathit{writes(s, x)}\} \right)
	\not\in \mvis{}^{-1}(t)
\end{gather*}
The $\mathit{writes}$ predicate holds for a transaction $t$ and an object $x$ iff $t$ writes to $x$. We use $\mathit{max}_R(S)$ to denote the maximum element in the set $S$ according to the total order $R$. That is, the element $u \in S$ such that $\forall v \in S \ldotp v = u \lor (v, u) \in R$. Furthermore, $\mvis{}^{-1}$ denotes the converse of the $\mvis{}$ relation. Thus, this formula holds iff there is some transaction $t$ that updates an object $x$ and is not aware of the most recent version of $x$ (as per \arb{}).

We define NLU as an isolation level that prohibits lost updates:
\begin{gather*}
	\textsc{NLU} \doteq \neg \textsc{LU}
\end{gather*}
This definition enforces that transactions are always aware of the values they overwrite in the arbitration order, thus preventing lost update situations in which a transaction's update is blindly overwritten. Intuitively, all histories allowed by NLU can be explained by some execution that is free of lost updates. Conversely, a history $H$ is not allowed by NLU iff all the executions that could have given rise to $H$ witness a lost-update.

Update Atomic disallows lost-updates if it implies NLU, i.e., if all histories allowed by UA can be explained by an execution free of lost-updates:
\[
	\textsc{UA} \implies \textsc{NLU}
\]

We can verify this property with Isolde with the synthesis problem \(\spec{\textsc{UA}}{\textsc{NLU}}\). For a scope of 10 transactions, with 5 operations per transaction, Isolde is unable to find such a history, which is consistent with the expectation that \textsc{UA} prevents lost-updates. We can sketch a proof that this holds for arbitrary history sizes: since \textsc{UA} forces a visibility edge between any two transactions that update an item in common, it necessarily forces each transaction $T$ to be aware of \emph{every} transaction $S$ which (1) precedes $T$ in the \arb{} order, and (2) writes an object in common with $T$ (this is because $\mvis{} \subseteq \mar{}$).

Isolde allows us to go a step further and verify an even stronger property: that UA and NLU are equivalent.
This would mean that UA reduces the set of histories of RA by exclusively excluding histories that witness lost-updates, or, in other words, that UA is the \emph{weakest} level that (1) is strongest than RA, and (2) prohibits lost-updates.
This is the ideal scenario, since usually the more permissive an isolation level is, the greater flexibility there is in implementing it, which, in theory, can lead to more performant implementations.

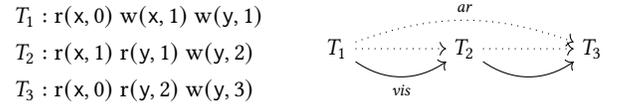
\begin{figure}[htb]
	\begin{center}
		\begin{minipage}[c]{.4\linewidth}
			\centering
			\begin{history}
				\txn{\r{x}{0} \w{x}{1} \w{y}{1}}
				\txn{\r{x}{1} \r{y}{1} \w{y}{2}}
				\txn{\r{x}{0} \r{y}{2} \w{y}{3}}
			\end{history}
		\end{minipage}
		\hspace{.5cm}
		\begin{minipage}[c]{.4\linewidth}
			\centering
			\begin{tikzpicture}
				\node[anchor=base] (t2) at (0, 0) {$T_1$};
				\node[anchor=base] (t1) at (0.5\textwidth, 0) {$T_2$};
				\node[anchor=base] (t3) at (\textwidth, 0) {$T_3$};
				\draw[arrows=->] (t2) to[out=-35, in=-145] node[below] {\scriptsize \vis{}} (t1);
				\draw[arrows=->] (t1) to[out=-35, in=-145] (t3);
				\draw[arrows=->, dotted] (t2) to[out=20, in=160] node[above] {\scriptsize \arb{}} (t3);
				\draw[arrows=->, dotted] (t2) -- (t1);
				\draw[arrows=->, dotted] (t1) -- (t3);
			\end{tikzpicture}
		\end{minipage}
	\end{center}
	\caption{A history allowed by \textsc{NLU} but not by \textsc{UA}.}
	\label{fig:nbu-not-ua}
\end{figure}

Since we have already verified one side of the implication, we just require verifying the other side:
\[
	\textsc{NLU} \implies \textsc{UA}
\]
This is done with the synthesis problem \(\spec{\textsc{UA}}{\textsc{NLU}}\). This time, Isolde is able to synthesize a history, which we show in Figure~\ref{fig:nbu-not-ua}, proving that \textsc{NLU} does not imply \textsc{UA}. This history is disallowed by \textsc{UA} since it contains two transactions ($T_1$ and $T_3$) that update the same data object ($y$) but cannot be related by \vis{}: both reads on $y$ by $T_2$ and $T_3$ force the visibility edges $\relatedBy{\mvis{}}{T_1}{T_2}$ and $\relatedBy{\mvis{}}{T_2}{T_3}$, which, in turn, establish the arbitration order $T_1 \rightarrow T_2 \rightarrow T_3$; this means that $T_1$ and $T_3$ could only be related by \vis{} through a $\relatedBy{\mvis{}}{T_1}{T_3}$ edge; however, since $T_3$ fails to read $T_1$'s write on $x$, this is not possible. On the other hand, the history is allowed by \textsc{NLU} because, as shown by the \vis{} and \arb{} relations in Figure~\ref{fig:nbu-not-ua}, it is possible to order the transactions in a way that every transaction sees the most recent version of the objects it overwrites.

The history synthesized by Isolde reveals the key difference between \textsc{UA} and \textsc{NLU}: while \textsc{UA} forces a visibility edge between any two transactions $T$ and $S$ that update an item $\mathtt{x}$ in common, such two transactions are only required to be related by \vis{} in \textsc{NLU} in the case that no other transaction ordered between $T$ and $S$ updates $\mathtt{x}$. Effectively, this means that a transaction $T$ that wrote to $x$ might not be aware of an older transaction $S$ that updated $x$ as long as it is aware of a more recent transaction than $S$ that also modified $x$. This is exactly the scenario produced by Isolde.

The counterexample produced by Isolde shows that \textsc{UA} is slightly over-restrictive when it comes to preventing the lost update anomaly since it rules out histories which do not witness any lost updates. If the goal is to prevent lost updates, it is acceptable for a transaction $T$ that writes to $x$ to \enquote{miss} an older version of $x$ as long as it is aware of one that supersedes it.

%% file: sections/technique.tex
\section{Synthesizing histories}\label{sec:technique}

In this section, we describe the history synthesis technique
implemented in Isolde.

For a more grounded presentation of our technique, we assume
isolation specifications are given in the framework of
\citet{biswasComplexityCheckingTransactional2019}. However, as
discussed in Section~\ref{sec:ir}, the same core technique applies to
any framework following the abstract execution paradigm introduced in
Section~\ref{sec:paradigm}.

\subsection{Challenges}

Recall that a history $H$ is allowed by a level $M$ if it can form a
valid execution under $M$ together with some commit-order
$\mathit{co}$. Conversely, a history is disallowed by a level $N$ if
there is \emph{no} commit-order $\mathit{co}$ which, together with
$H$, forms a valid execution under $N$. Thus, we formalize the
notation introduced in the previous section as follows:
\begin{gather*}
  \exists H \in \mathcal{S} \ldotp H \models \left\{ M_0, \ldots,
  M_m, \overline{N_0}, \ldots, \overline{N_n} \right\} \\
  \iff \\
  \exists H \in \mathcal{S} \ldotp \left(
    \bigwedge_{i=0}^{m} \left(\exists \mco \ldotp (H, \mco) \models
    M_i\right) \quad
    \land \quad
    \forall \mco \ldotp \left(\bigwedge_{i=0}^{n}(H, \mco)
    \not\models N_i\right)
  \right)
\end{gather*}

In practice, the (finite) universe of histories $\mathcal{S}$ used
for a synthesis problem is specified indirectly by providing the
following parameters that describe the format of the synthesized
history: the number of transactions, the number of operations
performed by each transaction, the set of database objects and values
to consider and, finally, the initial value of all objects in the
database. Note that, even for a few transactions,
operations, objects, and values, the finite set $\mathcal{S}$ of
possible histories is immense, and precludes any form of analysis by
explicit exhaustive search.

Our approach builds on top of SAT-based tools. This naturally follows
from the generality of our problem statement --- working with
arbitrary isolation level specifications.

The problem of synthesizing histories that are only required to be
allowed under a given set of isolation levels (i.e., $\exists H \in
\mathcal{S} \ldotp H \models \left\{ M_0, \ldots, M_m\right\}$) can
be directly encoded as a boolean satisfiability problem and solved by
an off-the-shelf SAT-solver. This requires translating (by
skolemization) the existentially quantified structures of a history
and $m$ binary relations on transactions (representing the valid
commit-orders for levels $M_0$ through $M_m$), as boolean variables,
and translating the specifications of $\left\{ M_0, \ldots,
M_m\right\}$ as constraints on these variables. This is a standard
approach when building isolation checkers --- tools that take
real-world database histories and verify if they are allowed under a
given isolation level~\cite{zhangViperFastSnapshot2023,
  biswasComplexityCheckingTransactional2019,
kingsburyGretchenOfflineSerializability2023}. The only difference is
that, in those cases, the history's content is already
pre-determined, so the SAT problem needs only to find satisfiable
values for the remaining structures --- in our case, for the
variables representing the commit-orders.

The main challenge in achieving our history synthesis goal is to
guarantee that synthesized histories are \emph{disallowed} by a
particular isolation level. Recall that, for a history $H$ to be
disallowed by a level $N$, the following must hold:
\[
  \forall \mco \ldotp \left(H, \mco\right) \not\models N
\]
Because of the universal quantification on the higher-order variable
\co{} (which ranges over binary relations on transactions), this
formula cannot be skolemized and directly encoded as a satisfiability problem.

\subsection{Synthesis algorithm}

In order to synthesize histories that are disallowed by a given
level, we propose Algorithm~\ref{algo:cegis-implementation} based on
a general program-synthesis technique known as
\emph{Counter\-example-Guided Inductive Synthesis}
(CEGIS)~\cite{solar-lezamaCombinatorialSketchingFinite2006}, which
effectively enables solving formulas with higher-order universal
quantifiers using first-order
solvers~\cite{milicevicAlloyGeneralpurposeHigherorder2019}. In short,
CEGIS can be used to synthesize a $P$ (usually a program) which
satisfies some specification $\phi$ for all possible higher-order
values $T$ (usually possible valuations for input variables):
\[
  \forall T \ldotp \phi (P, T)
\]
The algorithm uses two solvers in a loop to synthesize the desired
$P$. One of the solvers is used as a \emph{synthesizer} for producing
candidate values for $P$, and the other as a \emph{verifier} for
checking the correctness of $P$ against all possible inputs. The key
idea of the algorithm is to use the counterexamples produced by the
verifier for failing candidates as feedback for guiding future
candidate-synthesis steps.

We now provide a walkthrough of our adaptation of the CEGIS algorithm
for the particular case of synthesizing histories that are disallowed
by an isolation level $N$, presented in
Algorithm~\ref{algo:cegis-implementation}. It is rather trivial to
generalize this algorithm to the general problem of synthesizing
histories allowed by a set of levels and disallowed by another set of
levels. We describe only the restricted version of algorithm to make
the presentation easier to follow.

\revised{%
  \cleanorextended{%
    The typing discipline implicit in the algorithm and the
    definitions of helper functions such as \textsf{translate} and
    \textsf{encode} are given in the extended version of this
    paper~\cite{extended_paper}, which also includes proofs of the
    algorithm's soundness and (bounded) completeness.
  }{%
    The typing discipline implicit in the algorithm and the
    definitions of helper functions such as \textsf{translate} and
    \textsf{encode} are given in Appendix~\ref{app:algo}. The same
    Appendix also includes proofs of the algorithm's soundness and
    (bounded) completeness.
}}

\input{sections/algo/cegis}

Instead of a program $P$ that satisfies a given spec $\phi$ for every
possible set of inputs $T$, we look to synthesize a history $H$
which, together with \emph{any} relation \co{}, forms an invalid
execution under isolation level $N$.

We kick-start the algorithm by producing an initial candidate history
--- one which \emph{might} be disallowed by $N$. To do so, we look
for an execution $(H, \mco{})$ such that $(H, \mco) \not\models N$.
Recall that, if such execution exists, $H$ is not guaranteed to
violate $N$, since it might form a satisfying execution under $N$
together with a different commit-order relation.

To accomplish this first step, we start by encoding the problem's
scope --- which determines the size of histories to consider --- as a
set of boolean variables that model a database execution (line 2).
For example, to model the commit-order predicate, if $n$ is the
number of possible transactions in $\mathcal{S}$, we create $n \times
n$ boolean variables, with variable $\mco{}_{i,j}$ being true iff
$\relatedBy{\mco{}}{T_i}{T_j}$. The encoding of the remaining
predicates that model database executions can be done similarly.

The $\mathsf{translate}$ method takes in a first-order formula on
abstract executions and translates it to propositional logic over the
set of boolean variables created by $\mathsf{encode}$. This can be
done in a rather straightforward way, by unrolling all first-order
quantifiers and replacing predicate membership tests by the
respective boolean variable. In line 3, this is used to construct a
formula $I$ that searches for an execution containing the first
candidate history. Effectively, we look to solve the following formula:
\[
  \exists H, \mco{} \ldotp \left(H, \mco{}\right) \not\models N
\]
The actual search for a candidate is done in line 4. The
$\mathsf{solve}$ method takes in a formula in propositional logic,
converts it to CNF, and passes it as input to a SAT solver. In case
the formula is satisfiable, it returns a valuation of the boolean
variables for which the input formula holds. Otherwise, it returns $\bot$.

If the search for the first candidate returns $\bot$, meaning there
is no history under $\mathcal{S}$ disallowed by $N$, the algorithm
terminates unsuccessfully. Otherwise, $\mathit{cand}_0$ holds the
value of a concrete execution $(H_0, \mco{}_0)$ which violates $N$.
If this is the case, we proceed to verify if the candidate history
$H_0$ is in fact disallowed by $N$. This is done by looking for a
counterexample --- a witness \co{} relation which, together with $H$,
forms a valid execution under $N$:
\[
  \exists \mco{} \ldotp \left(H_0, \mco{}\right) \models N
\]
This is done by checking the satisfiability of the formula that
results from (1) translating $N$ to a propositional formula over the
boolean variables $B$ (line 7), and (2) substituting the variables
that model the history part of the execution by their corresponding
boolean value in the candidate valuation (line 8). Given a predicate
such as \co{}, $\mco{} \mapsto \mco_{m}$ is a shorthand notation to
denote the substitution of all variables that encode $\mco{}$ by the
respective value in the valuation $m$, that is $\mco{}_{i,f} \mapsto
m[\mco{}_{i,f}]$ for all possible transaction $i$ and $j$. Notation
$H \mapsto H_m$ is a shorthand notation for substituting all
first-order predicates that encode history $H$ by the respective valuation.

If the verification problem returns $\bot$, the process finishes
successfully with the most recent candidate as the synthesized
history. Otherwise, the counterexample --- the \co{} relation that
makes the candidate history valid under $N$ --- is used to guide the
search for new candidates. This is done by excluding from future
candidate-searching steps histories that would be refuted by the same
counterexample. That is, after each failing candidate
$\mathit{cand}_i$, refuted by counterexample $\mathit{cex}_i$, we
extend the candidate-searching formula (line 12) with new clauses
that prevent future candidate histories from forming valid executions
under $N$ with the concrete value of \co{} found in $\mathit{cex}_i$
(line 10). Effectively, the formula being solved at each
candidate-synthesis step is the following:
\[
  \exists H, \mco \ldotp \left(H, \mco{}\right) \not\models N
  \bigwedge_{j=0}^{i} \left(H, \mco_{cex_j}\right) \not\models N
\]
This process is repeated until either (1) the candidate-searching
problem becomes unsatisfiable (line 13), in which case no history is
synthesized, or (2) the search for a counterexample (line 15) fails,
in which case the most recent candidate is returned as the
synthesized history (line 17).

Extending this algorithm to work for the general problem of producing
a history $H \in \mathcal{S}$ allowed by levels $\left\{ M_0, \ldots,
M_m \right\}$ and disallowed by levels $\left\{N_0, \ldots, N_n
\right\}$ is straightforward. For each level $M_i$, we introduce a
new set of boolean variables representing a binary relation
$\mco{}_i$, each of which we enforce to form, together with $H$, a
valid execution under $M_i$. Furthermore, dealing with the
requirement of disallowing multiple levels requires no modification
to the algorithm itself, since the problem of synthesizing a history
$H$ that is disallowed under multiple levels $\left\{ N_0, \ldots,
N_n \right\}$ can be reduced to synthesizing a history disallowed by
an isolation level defined as $\bigvee_{i=0}^{n} N_i$.

\revised{
  \subsubsection{Mixing Isolation Frameworks}\label{sec:ir}
  Although we presented the algorithm in the light of the framework
  of \citet{biswasComplexityCheckingTransactional2019}, the same core
  technique applies to any of the abstract execution frameworks discussed in
  Section~\ref{sec:background}. Because these frameworks agree on the same
  notion of a history, adapting the algorithm to another framework
  essentially requires changing the boolean encoding of the witness
  execution structures. For instance, when working with the framework
  of \citet{ceroneFrameworkTransactionalConsistency2015}, instead of
  a single binary relation \co{}, the SAT-solving problems need to
  search for two binary relations (representing the \vis{} and \arb{}
  relations that are used to model executions in this formalism).

  Furthermore, the shared abstract execution paradigm allows us to
  mix isolation definitions expressed in different frameworks in the
  same synthesis problem. As explained above, each level $M_i$ in
  $\left\{ M_0, \ldots, M_m \right\}$ requires extending the SAT
  problems with a set of boolean variables that, together with the
  synthesized history, model a database execution allowed under
  $M_i$. Since, besides the history part, these are all independent
  executions, these sets of variables can encode different types of
  witness executions. In candidate-searching SAT problems, the same
  applies to $\left\{N_0, \ldots, N_n \right\}$ except that, as
  explained above, levels defined under the same framework can be
  merged onto a single constraint. When verifying a candidate, it needs to
  be checked for every level in $\left\{N_0, \ldots, N_n \right\}$.
  To do this, we use a separate SAT problem for each level $N_i$. If
  all checking problems are $\mathsf{UNSAT}$, the candidate is
  returned as the synthesis solution; otherwise, we use the
  counterexamples produced in each $\mathsf{SAT}$ problem as feedback
  in the next candidate-searching step as described above.

  The abstract execution paradigm effectively allows us to treat
  histories as an \emph{intermediate
  representation} (IR) that is meaningful across frameworks. This
  enables fully automated (bounded) comparison of definitions given
  in different frameworks. While the abstract execution paradigm has
  been informally
  identified in prior work~\cite{szekeresMakingConsistencyMore2018a},
  Isolde is, to our
  knowledge, the first tool to \emph{operationalise} it algorithmically.
}


\subsection{Implementation}\label{ssec:optimizations}

Instead of directly implementing
Algorithm~\ref{algo:cegis-implementation} using a SAT solver, Isolde
uses Kodkod~\cite{torlakKodkodRelationalModel2007}, a relational
model finder that can directly be used to solve first-order problems
using off-the-shelf SAT solvers. Essentially, Kodkod simplifies the
implementation of the $\mathsf{encode}$ and $\mathsf{translate}$
methods of our algorithm. As was showcased in the previous section,
Kodkod has the added benefit of offering a high-level API for
constructing relational formulas and expressions, which makes Isolde
easily extendable with custom isolation specifications.

The remainder of this section discusses additional implementation
decisions we explored with the goal of speeding up synthesis times.
Their actual impact to the final performance is evaluated in
Section~\ref{sec:evaluation}.

\subsubsection{History encoding}\label{sssec:encoding}
Each candidate-synthesis step requires finding suitable values for a
set of boolean variables representing a history. A direct encoding of
the classical notion of a history requires representing the linear
sequence of operations associated with each transaction. Instead, to
produce more efficient SAT problems, we abstract away the internal
operations performed by transactions and use what we call
\emph{abstract transactions}. Rather than a total order of
operations, an abstract transaction is directly associated with its
set of external reads and its set of final writes. An external read
is a read operation on an object that precedes any read or write to
that object. A final write is a write operation that is not followed
by any other write to the same object. This representation prevents
us from having to encode the concept of operations and all the
relations associated with it (e.g., the total order of operations
associated with each transaction). Instead, abstract transactions are
directly associated with their external operations by means of two
\emph{ternary} first-order predicates (encoded by the respective
boolean variables), both of which associate transactions, objects and
values. These predicates, called \texttt{reads} and \texttt{writes},
represent, respectively, the set of all external reads, and the set
of all final writes. As an example, the presence of the tuple $(t, x,
n)$ in the \texttt{reads} predicate would indicate that transaction
$t$ does an external read of value $n$ from object $x$.

This makes certain histories impossible to represent, e.g., a history
in which a transaction fails to read a value written by itself.
However, all isolation levels we are aware of guarantee that (1) a
transaction's read operations are consistent with the results of its
previous operations (a property often denoted as \emph{internal
consistency}), and (2) only the last write to each object by each a
transaction is visible to other
transactions~\cite{biswasComplexityCheckingTransactional2019}. Most
isolation levels used in practice constrain only the final writes and
external reads of the transactions in a history. Thus, our
alternative transaction encoding has little impact on expressiveness.

\subsubsection{Fixing the commit-order}
In most isolation frameworks, an abstract execution includes a total
order on the transactions of its history representing the
commit-order. In order to speed up candidate searches, we
preemptively choose an arbitrary total order over transaction
identifiers and encode it as the commit-order. This allows us to
(1) simplify the formula passed into the solver by removing all the
clauses used to enforce the total ordering constraint of this
relation, and (2) reduce the universe of possible candidates by
removing redundant executions from the search space. \revised{Note
  that this does not exclude any valid candidate from the
  search space, since the mapping from transactions to their
  operations is still completely left to
the solving step.}

Applying this optimization in the algorithm shown above consists in
substituting, in the formula generated by the $\mathsf{translate}$
method, variable $\mco{}_{i,j}$ by $\mathit{true}$ when $i < j$ and
by $\mathit{false}$ otherwise.

Note that, when synthesizing histories that must satisfy more than
one isolation level, this optimization can only be performed for one
of these levels. This is because we cannot pre-determine how every
other execution relates to the one whose commit-order relation got
fixed. 

\subsubsection{Smart candidate search}
Algorithm~\ref{algo:cegis-implementation} shows how to synthesize a
history disallowed by a level $N$. Most often, we also want the
synthesized history to be allowed under a given level $M$. This
requires restricting the candidate-searching step to only consider
histories that satisfy $M$. This could be done by modeling an extra
\co{} relation and enforcing it to form, together with the
synthesized history, an execution allowed by $M$:
\[
  \left(\exists \mco \ldotp (H, \mco{}) \not\models N\right)
  \quad \land \quad
  \left(\exists \mco \ldotp (H, \mco{}) \models M\right)
\]
We, however, use a simpler formulation that prevents adding any
additional variables to the problem:
\[
  \exists \mco \ldotp (H, \mco{}) \not\models N \land (H, \mco{}) \models M
\]
Instead of an independent existential constraint forcing $H$ and some
\co{} to satisfy $M$, we merge this in the existential constraints
that searches for an execution that violates $N$. This formulation is
correct since it does not exclude from the search space viable
candidate histories. \revised{Proving it is trivial: for any
  candidate $H$, we must find a binary relation \co which, together
  with $H$, forms a valid execution under $M$; at the same time, $H$ is
  required to violate $N$ for every possible commit-order relation;
  therefore, any valid candidate must form, with some commit-order, an
  execution that simultaneously satisfies $M$ and violates $N$. We can
  thus conclude that we do not miss any valid candidate by searching
for candidate executions that both satisfy $M$ and violate $N$.}



%% file: sections/algo/cegis.tex
\begin{algorithm}[tb]
  \caption{\revisedinline{Synthesizing a history $H \in \mathcal{S}$
  disallowed by $N$}}
  \label{algo:cegis-implementation}
  \begin{algorithmic}[1]
    \Ensure \revisedinline{$H \neq \bot \Leftrightarrow H \in \mathcal{S}
    \land \forall \mco{} \ldotp (H, \mco{}) \not\models N$}
    \State \revisedinline{$i \gets 0$}
    \State \revisedinline{$B \gets \mathsf{encode}\left(\mathcal{S}\right)$}
    \State \revisedinline{$I \gets \mathsf{translate}\left(B,\neg N\right)$}
    \State \revisedinline{$\mathit{cand}_0 \gets \mathsf{solve}\left(I\right)$}
    \If{\revisedinline{$\mathit{cand}_0 \neq \bot$}}
    \State \revisedinline{$C \gets \mathsf{translate}\left(B, N\right)$}
    \State \revisedinline{$\mathit{cex}_0 \gets \mathsf{solve}\left(C \left[H
    \mapsto H_{\mathit{cand}_0}\right] \right)$}
    \While{\revisedinline{$\mathit{cex}_0 \neq \bot$}}
    \State \revisedinline{$L_i \gets \mathsf{translate}\left(B, \neg N\right)
    \left[\mco{} \mapsto \mco_{\mathit{cex}_i}\right]$}
    \State \revisedinline{$i \gets i + 1$}
    \State \revisedinline{$\mathit{cand}_i \gets \mathsf{solve}\left(I
    \bigwedge_{j=0}^{i-1} L_j\right)$}
    \If{\revisedinline{$\mathit{cand}_i = \bot$}} \Break \EndIf
    \State \revisedinline{$\mathit{cex}_i \gets \mathsf{solve}\left(C \left[H
    \mapsto H_{\mathit{cand}_i}\right] \right)$}
    \EndWhile
    \EndIf
    \State \revisedinline{$H \gets \mathit{cand}_i$}
  \end{algorithmic}
\end{algorithm}

%% file: sections/evaluation.tex
\revised{%
\section{Evaluation}\label{sec:evaluation}}

\def\mathdefault#1{#1} 

\revised{%
  Our evaluation aimed to answer the following research questions:
  \begin{enumerate}[label=\textbf{RQ\arabic*:}]
    \item What is the impact of using CEGIS compared with simpler
      enumeration strategies?
    \item How impactful are the implementation decisions of fixing the
      commit-order and using smart candidate search?
    \item Can Isolde efficiently solve synthesis problems for meaningful scopes?
  \end{enumerate}

  Our research questions focus on Isolde's performance. As previously
  mentioned, the practical value of Isolde depends on its performance.
  A tool that takes hours to produce a single counterexample cannot
  serve as a genuine aid to the specification process. For Isolde to be
  useful --- enabling researchers to quickly test candidate
  definitions, obtain concrete counterexamples, and iterate before
  investing in a full formal proof --- synthesis must be fast.

  The first two questions address the impact of the implementation
  decisions behind Isolde. For RQ1, in order to assess the impact of
  using CEGIS, we have developed two baseline synthesis implementations
  that do not rely on any learning strategy to speed up convergence to
  a solution. The first one is a brute-force approach that
  programmatically iterates through all possible histories under a
  given scope, and stops once either (1) it finds a history that
  satisfies the synthesis problem, or (2) it exhausts the search space.
  To verify that a history $H$ is a solution for a problem $\spec{M_0,
  \ldots, M_m}{N}$, it iterates through all possible abstract
  executions containing $H$ and asserts, using a SAT-solver, that (1)
  each definition $M$ holds in at least one execution, and (2) $N$ does
  not hold in any execution. Our second baseline uses SAT-solving for
  both synthesizing candidates, and verifying them. The main difference
  to the CEGIS approach is that it does not use the counterexamples
  generated in verification as feedback to the synthesis. Instead, we
  encode the first-order constraints $M_0, \ldots, M_m, \neg N$ as
  propositional logic and pass it to a SAT-solver, and iterate through
  the solutions, verifying each one for $N$ by using an additional
  solver. The process ends when either the verifying solver is not able
  to find an execution satisfying $N$, or the synthesis solver runs out
  of solutions. Like the main implementation Isolde, we used Kodkod for
  ease of translation between first-order and propositional logic.

  For RQ2 we conducted an ablation study to assess the impact of fixing
  the commit-order and using smart candidate search, because these were
  the two implementation decisions whose positive impact on performance
  was not clear upfront. Using a more abstract history encoding
  substantially reduces the number of variables in the constraint
  problems, which causes a significant performance improvement.

  With RQ3, we aim to assess Isolde's usefulness in practice. The scope
  used in a synthesis problem is important because it informs our
  confidence in the UNSAT results produced by Isolde: the larger the
  scope is, the more representative is the set of histories considered,
  and the safer it is to extrapolate Isolde's output to the unbounded
  set of \textit{all} histories. However, larger scopes result in
  computationally harder constraint solving problems and, consequently,
  longer solving times. By \enquote{meaningful scope} we mean one that
  can give a good approximation on the unbounded results for a given
  specification. We base our notion of a meaningful scope on the work
  of \citet{cuiUnderstandingTransactionBugs2024}, in which they argue,
  based on large collection of real-world transaction histories, that
  transaction anomalies follow the small-scope
  hypothesis~\cite{jacksonElementsStyleAnalyzing1996}, i.e., can be
  triggered with a small number of simple transactions operating on
  just a few data items. More specifically, all the transaction bugs
  they analyzed could be reproduced with up to five transactions (with
  most requiring only three transactions); most anomalies could be
  reproduced using no more than four SQL statements per transaction;
  and most anomalies required only one database table with initially no
  more than five rows of data. While they focus on actual SQL
  transactions operating on real databases, the same general conclusion
  can be applied to our conceptual database model which is restricted
  to reads and writes over key-value stores. Based on these
  conclusions, we use \enquote{meaningful scope} to refer to any scope
  with at least five transactions, five different objects, and five values.

  \subsection{Benchmark}

  We now characterize the set of synthesis problems that served as the
  benchmark for our evaluation.

  To obtain a varied and representative benchmark, we considered the
  two following orthogonal properties of synthesis problems: whether
  they are SAT or UNSAT, and whether they involve definitions given in
  the same formal frameworks or in different frameworks. We expect both
  of these to have an impact on performance: UNSAT problems should take
  longer than SAT ones, because they require Isolde to fully exhaust
  the set of potential candidates before concluding that no satisfying
  history exists; and problems involving multiple frameworks should
  take more than those involving a single framework, because they
  always require encoding multiple execution structures --- this
  requires more SAT clauses, which is the main factor impacting
  solver performance.

  For SAT problems, we considered the following problems:
  \begin{itemize}
    \item One synthesis problem for each of the edges of the isolation
      level hierarchy showed in Figure~\ref{fig:isolation-levels}.
      Namely, for each direct implication \(A \implies B \), our
      benchmark includes the problem
      of synthesizing a history allowed by B but not by A (i.e.,
      \(\spec{B}{A}\)).
    \item To also include problems that require histories to satisfy
      more than one level, our benchmark includes the problem of
      synthesizing the anomaly presented in Section~\ref{ssec:fekete_anomaly}.
    \item The SAT problem we used in the previous section to prove that
      Plume's definition of RA is not equivalent to its axiomatic counterpart.
  \end{itemize}
}

\begin{figure}[t]
  \centering
  \[
    \textsc{Ser} \implies \textsc{SI} \implies \textsc{PC} \implies
    \textsc{CC} \implies \textsc{RA}
  \]
  \caption{A hierarchy of five isolation levels that have been
  formalized in both frameworks considered.}
  \label{fig:isolation-levels}
\end{figure}

\revised{%
  Each of these problems (except the ones related to Plume's
  definitions) was instantiated using every combination of the two
  frameworks considered. For instance, the specification
  \(\spec{\textsc{RA}}{\textsc{CC}}\) gives rise to a total of four
  problems (using subscripts $\mathcal{B}$ and $\mathcal{C}$ to
    distinguish definitions in the frameworks of
    \citeauthor{biswasComplexityCheckingTransactional2019} and
  \citeauthor{ceroneFrameworkTransactionalConsistency2015} respectively):
  \(\spec{\textsc{RA}_{\mathcal{B}}}{\textsc{CC}_{\mathcal{B}}}\),
  \(\spec{\textsc{RA}_{\mathcal{C}}}{\textsc{CC}_{\mathcal{C}}}\),
  \(\spec{\textsc{RA}_{\mathcal{B}}}{\textsc{CC}_{\mathcal{C}}}\), and
  \(\spec{\textsc{RA}_{\mathcal{C}}}{\textsc{CC}_{\mathcal{B}}}\).

  For UNSAT problems involving specifications of different frameworks,
  we considered the problems involved in verifying definition
  equivalence for each of the levels in
  Figure~\ref{fig:isolation-levels}. For instance, for SER, we include
  problems
  \(\spec{\textsc{Ser}_{\mathcal{B}}}{\textsc{Ser}_{\mathcal{C}}}\) and
  \(\spec{\textsc{Ser}_{\mathcal{C}}}{\textsc{Ser}_{\mathcal{B}}}\).

  For UNSAT problems involving specifications in the same framework, we
  attempted to find counterexamples for each of the implications of
  Figure~\ref{fig:isolation-levels}. For example, for the implication
  \(\textsc{Ser} \implies \textsc{SI}\) we included the UNSAT problem
  of finding a history allowed by \textsc{SER} and not by \textsc{SI}.
  Each direct implication gives rise to two concrete problems, each for
  a particular framework. Furthermore, for this category of problems,
  we also considered the three UNSAT problems involved in verifying the
  isolation definitions of Plume discussed in
  Section~\ref{sec:equivalenceofspecs}.

  Table~\ref{tab:benchmark} summarizes the contents of the benchmark
  used in our evaluation. For all problems, we considered scopes with
  five objects, and five values, with the number of transactions
  ranging from three to ten.
}

\begin{table}[tb]
  \centering
  \begin{tabular}{lcc}
    & SAT & UNSAT \\
    \hline
    Single framework    & 11  & 11    \\
    Multiple frameworks & 14  & 10    \\
  \end{tabular}
  \caption{Number of problems of each type in the benchmark.}
  \label{tab:benchmark}
\end{table}

\begin{figure}[t]
  \centering
  \scalebox{.55}{\input{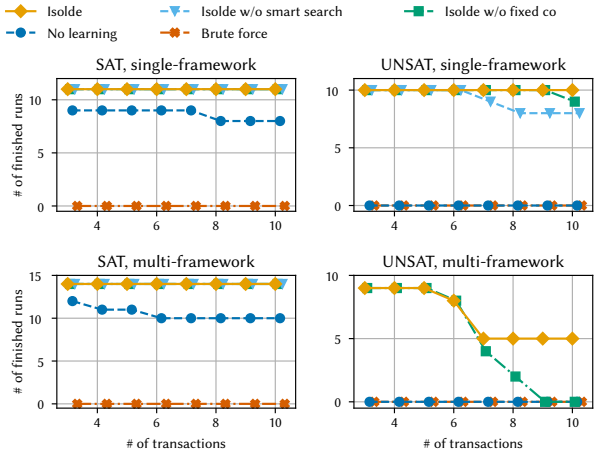}}
  \caption{\revisedinline{Total instances solved within the one-hour
      timeout across
  different implementations, scopes and problem types.}}
  \label{fig:plot1}
\end{figure}
\begin{figure}[t]
  \centering
  \scalebox{.45}{\input{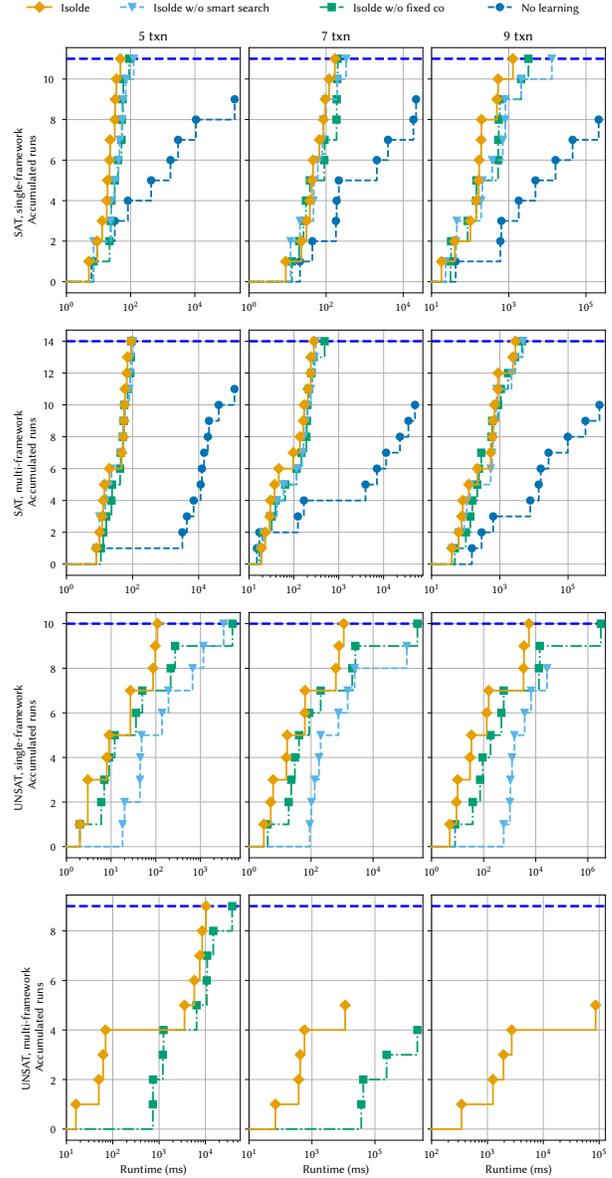}}
  \caption{\revisedinline{Runtime performance comparison across different
  implementations, scopes, and problem types.}}
  \label{fig:cactus-times}
\end{figure}

\revised{%
  \subsection{Results and Discussion}

  Isolde was ran for all problems discussed in the benchmark above. In
  order to address RQ1, we ran the same benchmark for the two baseline
  implementations discussed. To address RQ2, we additionally ran two
  versions of Isolde, each with a different optimization (smart
  candidate search, or fixed commit-ordered) disabled. Runs were timed
  out after one hour.
  The UNSAT problems $\spec{RA_{\mathcal{B}}}{RA_{\mathcal{C}}}$ and
  $\spec{PC_{\mathcal{C}}}{RA_{\mathcal{C}}}$ could be solved without
  any SAT-solving, particularly during Kodkod's translation phase. This
  happens because Cerone's framework is targetted at levels which
  guarantee Read Atomic, so the constraint $\neg \textsc{RA}$
  introduces a contradiciton that is caught during translation to
  propositional logic. Since these measurements are not representative
  of real problems, we excluded them from our results.

  Our experimental results are summarized in Figure~\ref{fig:plot1} and
  Figure~\ref{fig:cactus-times}. Figure~\ref{fig:plot1} shows the number of
  problems that were solved under the one hour timeout for each of the
  implementations considered, for each of the classes of problems we
  considered. Figure~\ref{fig:cactus-times} shows, for each class of problem,
  and for three of the scopes considered, how different
  implementations compare in runtime. In each sub-plot, the x-axis represents
  runtime (in milliseconds, on a logarithmic scale) and the y-axis
  represents the number of benchmark instances that completed within
  that runtime. Each curve corresponds to a particular implementation:
  a point $(t, k)$ on a curve means that exactly $k$ instances produced
  a result in at most $t$ milliseconds. The blue horizontal line in
  each sub-plot denotes the total number of problems for that
  problem-class in our benchmark. Because the brute-force
  implementation fails to produce any results for the considered
scopes, we exclude it from Figure~\ref{fig:cactus-times}.}

\revised{%
  \paragraph{RQ1} As showcased in Figure~\ref{fig:plot1}, the
  brute-force implementation failed to finish for even the minimum
  scope considered. This goes to show the tremendous rate at which the
  number of possible histories grows with scope size.
  As an experiment, we left this implementation running overnight for
  more than twelve hours for the problem
  $\spec{RA_{\mathcal{B}}}{\textsc{Ser}_{\mathcal{B}}}$ with a scope of
  three transactions, three values, and three objects, and it did not
  finish under that time limit.

  In terms of the number of completed problems under the timeout-limit,
  the solver-based baseline improves upon the brute-force baseline.
  However, it times out on all UNSAT problems. This supports the
  positive impact of using CEGIS ---
  it allows solving problems otherwise untractable with enumeration
  techniques that do not incorporate any learning.
  Furthermore, as shown in Figure~\ref{fig:cactus-times}, for the SAT
  problems that the baseline is able to solve, its runtime is generally
  worse than Isolde's.

  \paragraph{RQ2} As shown in Figure~\ref{fig:cactus-times}, the impact of
  both optimizations is most felt in UNSAT problems. For
  single-framework UNSAT problems, removing either optimization has a
  noticeable negative impact on performance, namely causing some
  executions that otherwise complete under one hour to time-out under
  this limit. A similar effect is observed when disabling the fixed
  commit-order optimization for multi-framework UNSAT problems to an
  even greater degree (e.g., without this optimization, Isolde times
  out for all problems under this category for more than eight transactions).

  Note that smart candidate search optimization is not applicable on
  multi-framework problems. This is because, in these problems, the
  positive and negative isolation definitions belong to different
  frameworks and therefore require independent witness execution
  structures; as a result, the two existential quantifiers in the
  candidate-searching step cannot be merged, and the optimization has
  no effect in practice.

  \paragraph{RQ3} Figure~\ref{fig:cactus-times} demonstrates that Isolde is
  capable of solving synthesis problems at meaningful scopes --- those
  with at least five transactions, five objects, and five values ---
  within practical time limits. For SAT problems, both single- and
  multi-framework instances are resolved in at most a few (<10) seconds
  across all scopes considered. UNSAT problems are inherently harder and
  runtimes are correspondingly higher. Nevertheless, despite noticeable
  performance degradation with more than five transactions in the case
  of multi-framework UNSAT problems, Isolde successfully solves all
  UNSAT problems within the one-hour timeout for scopes of five
  transactions, five objects, and five values. These results confirm
  that Isolde operates at a scale sufficient to give meaningful
  confidence in its outputs, supporting its practical utility as a
lightweight verification aid during the specification design process.}

\cleanorextended{%
  \revised{The full version of this paper~\cite{extended_paper} provides
    complementary analysis of CEGIS candidate counts and SAT formula
    sizes by showing how these metrics scale with transaction scope
  across implementations and problem types.}
}{%
  \revised{Appendix~\ref{app:eval} provides complementary analysis:
    Figures~\ref{fig:cactus-cand} and~\ref{fig:cactus-clauses}
    decompose solver performance into CEGIS candidate counts and SAT
    formula sizes, isolating the contribution of learning-based search
    guidance from encoding complexity;
    Figures~\ref{fig:compare-metrics-impl}
    and~\ref{fig:compare-metrics-ptypes} show how these metrics scale
  with transaction scope across implementations and problem types.}
}

All results shown in this section were obtained with a machine with
\SI{32}{\gibi\byte} of RAM, and an AMD Ryzen 7 PRO 6850U CPU, with a
\verb|x86_64| architecture, running Fedora Linux 40, with kernel
version 6.14.5. Isolde was configured to use Glucose as the backend
SAT-solver for Kodkod. Experiments were also conducted using Minisat
(which performed similarly), and SAT4j (which exhibited considerably
worse performance). All problems were run multiple times for each
scope and the tables and figures report the average times.

%% file: sections/relatedwork.tex
\section{Related Work}\label{sec:related-work}
%
%
%
Leveraging formal definitions of isolation to build software that
helps deal with the inherent difficulty of using weak isolation
levels is a prevalent line of research. Here, we focus on two
relevant types of applications.

\paragraph{Isolation checkers}
Isolation checkers verify the isolation guarantees offered by
real-world databases. The most prevalent approach is opaque-box
checking: workloads are injected into the system under test, and
the returned output is analyzed for patterns that reveal isolation
violations~\cite{zhangViperFastSnapshot2023,
  kingsburyGretchenOfflineSerializability2023, tanCobraMakingTransactional2020,
  huangEfficientBlackBoxChecking2023, kingsburyElleConsistencyModel,
biswasComplexityCheckingTransactional2019, liLeopardBlackBoxApproach2023}.
Because of the extensive concurrent
workloads used to increase the chance of triggering anomalies,
research has largely focused on designing efficient
verification algorithms for long histories --- a different regime
from Isolde's SAT-based checking, which is general-purpose but
targets small scopes.

\paragraph{Verifying application correctness under weak isolation}
Research and development on distributed database systems on a
planetary-scale is divided according to the CAP theorem
\cite{gilbertBrewerConjectureFeasibility2002} between ensuring the
conditions needed for strict serializability
\cite{corbettSpannerGoogleGlobally2013} or devising weaker isolation
levels that are still useful to application developers, but allow
highly available and efficient implementations
\cite{DeCandia2007-ga,Lloyd2011-tq,Sovran2011-hv,S_Elnikety_undated-gv,Ardekani2013-og,Lloyd2013-di,Bailis2016-sp,Akkoorath2016-nb,Spirovska2019-ka,Liu2024-cc}.
Taking advantage of the latter leads to the development of tools to
verify the correctness of applications running on top of weakly
consistent storage
systems~\cite{biswasMonkeyDBEffectivelyTesting2021,
  ganIsoDiffDebuggingAnomalies2020,
  nagarAutomatedDetectionSerializability2018,
  rahmaniCLOTHODirectedTest2019,
warszawskiACIDRainConcurrencyRelatedAttacks2017}. These tools look to
help developers detect undesirable behavior of their application
caused by non-serializable executions of the underlying storage system.
These tools often leverage custom isolation properties to support
algorithmic optimizations. Isolde can help in the design and
verification of these properties.

\revised{%
  \paragraph{LLM-assisted formal specification} An emerging use case
  for tools like Isolde is as a backend validator in LLM-assisted
  specification workflows. Large language models have shown promise
  in generating formal specifications, but offer no correctness
  guarantees. Isolde can serve as an automated oracle in this
  setting: given an LLM-generated specification, it either confirms
  equivalence with a reference definition up to a given bound, or
  produces a concrete counterexample. Crucially, these
  counterexamples can be fed back to the LLM as evidence of the
  specification's flaw, enabling an iterative refinement loop that
  combines the generative power of language models with the formal
  rigor of automated synthesis. Evaluating this workflow is beyond
  the scope of this paper, but Isolde's design readily supports it.
}

%% file: sections/conclusion.tex
\section{Conclusion}\label{sec:conclusion}

In this work we presented a technique for automatically synthesizing
transactional histories allowed by a set of isolation levels and
disallowed by another, implemented in a tool called Isolde. Our
approach is framework-agnostic: it applies to any isolation formalism
that follows the \emph{abstract execution}
paradigm~\cite{szekeresMakingConsistencyMore2018a}, which covers most
theoretical frameworks in the
literature~\cite{adyaGeneralizedIsolationLevel2000,
  ceroneFrameworkTransactionalConsistency2015,
  crooksSeeingBelievingClientCentric2017,
biswasComplexityCheckingTransactional2019}. As a result, Isolde can
be used to compare isolation levels specified in different formalisms.

Isolde shines as a verification tool for researchers that work on
novel formal results about isolation levels. We have used it to
discovered a specification
bug in a modern isolation checker, and to reproduce
a famously elusive result about Snapshot Isolation.

%% file: sections/appendix/appendix-algo.tex
\section{Formalization of Algorithm 1}\label{app:algo}

This appendix formalizes the types and semantic objects implicit in
Algorithm~\ref{algo:cegis-implementation} of
Section~\ref{sec:technique}, that implements the core Isolde's
history synthesis procedure. This procedure  --- henceforth
\textsf{synth} --- takes a scope and an isolation level specification
expressed in first-order logic, and produces a solution:
\[
  \mathsf{synth} : \mathsf{Scope} \times \mathsf{FolFormula} \to
  \mathsf{Solution}
\]
The types \textsf{Scope} and \textsf{FolFormula} will be explained in
the following sections. A \textsf{Solution} represents the result
produced by a SAT solver. It is either SAT or UNSAT. If SAT, it
contains a mapping from propositional variables to boolean values. In
case of \textsf{synth}, this mapping represents a history that solves
the given synthesis problem.
\begin{align*}
  \mathsf{Solution} & \coloneq \{\bot\} \cup \mathsf{Instance} \\
  \mathsf{Instance} & \coloneq \mathsf{BoolVar} \to \{T, F\}
\end{align*}

\subsection*{Constraint language}

Isolation constraints are written in a many-sorted first-order logic
(FOL) language, with three disjoint sorts: \textsf{Txn},
\textsf{Obj}, and \textsf{Val}. These represent, respectively, the
set of transactions in a history, the set of all database objects,
and the set of all values. We use the type \textsf{FolFormula} to
refer to formulas written in this language.

Here, we model abstract executions using the encoding explained in
Section~\ref{sssec:encoding}: the operations of the transactions in a
history are modelled by two ternary first-order predicates, called
\textsf{reads} and \textsf{writes}. Additionally, the \textsf{so}
predicate models the session-order, and predicate \textsf{co} the
commit-order. Thus, our contraint language contains the following
predicate symbols:
\begin{align*}
  \mathsf{writes} & : \mathsf{Txn} \times \mathsf{Obj} \times \mathsf{Val}, \\
  \mathsf{reads}  & : \mathsf{Txn} \times \mathsf{Obj} \times \mathsf{Val}, \\
  \mathsf{so}     & : \mathsf{Txn} \times \mathsf{Txn},                     \\
  \mathsf{co}     & : \mathsf{Txn} \times \mathsf{Txn}.
\end{align*}

We use the following metavariables to describe different categories of syntax:
\[
  \begin{array}{rl}
    \varphi       & \text{formulas}        \\
    \tau          & \text{sorts}           \\
    t             & \text{terms}           \\
    x             & \text{variables}       \\
    c             & \text{constants} \\
    P             & \text{predicates}
  \end{array}
\]

Next we formalize the abstract syntax for the considered FOL fragment.
\[
  \begin{array}{ccl}
    t       & ::= & x \mid c
    \\
    \tau    & ::= & \mathsf{Txn} \mid \mathsf{Obj} \mid \mathsf{Val}
    \\
    \circ   & ::= & \land \mid \lor \mid \; \Rightarrow
    \\
    P       & ::= & \mathsf{writes} \mid \mathsf{reads} \mid
    \mathsf{so} \mid \mathsf{co} \\
    \varphi & ::= & P(t, \ldots, t)
    \mid \neg \varphi
    \mid \varphi \, \circ \,  \varphi
    \mid \forall x{:}\tau.\, \varphi
    \mid \exists x{:}\tau.\, \varphi
  \end{array}
\]

Constants correspond correspond to particular members of the sorts in
question. While they are generally not used in the specifications
themselves, their inclusion in the grammar helps define the
translation from FOL to propositional logic, which is shown later.

\subsection*{Scope encoding}

When solving a synthesis problem, we assume the domains of all three
sorts are finite and fixed by the input scope. A scope is a tuple
with three numbers that respectively bound the size of the sorts:
\[
  \mathsf{Scope} \coloneq \mathbb{N} \times \mathbb{N} \times \mathbb{N}
\]
A problem's scope determines the set of constants each sort
corresponds to. Given a scope \(\mathcal{S} = (\mathsf{txn},
\mathsf{obj}, \mathsf{val})\), we refer to the corresponding finite
sorts as \(\mathsf{Txn}_\mathcal{S}\), \(\mathsf{Obj}_\mathcal{S}\),
\(\mathsf{Val}_\mathcal{S}\), and define them as follows:
\begin{align*}
  \mathsf{Txn} &= \{t_0, \ldots, t_{\mathsf{txn} - 1}\} \\
  \mathsf{Obj} &= \{x_0, \ldots, x_{\mathsf{obj} - 1}\} \\
  \mathsf{Val} &= \{n_0, \ldots, n_{\mathsf{val} - 1}\}
\end{align*}

The \textsf{encode} function translates a scope \(\mathcal{S} =
(\mathsf{txn}, \mathsf{obj}, \mathsf{val})\) to the corresponding
finite sets and propositional variables that can be used to represent
an execution with $\mathsf{txn}$ transactions, $\mathsf{obj}$
objects, and $\mathsf{val}$ values.
\[
  \mathsf{encode} : \mathsf{Scope} \to \mathsf{Vars}
\]
Each of the three predicates gets encoded as set of propositional
variables. Namely, for each predicate, we use one variable for each
unique tuple that the predicate might contain, denoting the presence
(or absence) of that tuple in the predicate.

The return value of the \textsf{encode} function is a tuple:
\[
  \mathsf{Vars} \coloneq
  \Sigma\, S \in \text{Scope} .\;
  \begin{aligned}
    &\mathcal{P}(\textsf{Txn}) \times \mathcal{P}(\textsf{Obj})
    \times \mathcal{P}(\textsf{Val}) \;\times \\
    &\left(\mathsf{Txn}_\mathcal{S} \times \mathsf{Obj}_\mathcal{S}
    \times \mathsf{Val}_\mathcal{S} \to \mathsf{BoolVar} \right) \; \times \\
    &\left(\mathsf{Txn}_\mathcal{S} \times \mathsf{Obj}_\mathcal{S}
    \times \mathsf{Val}_\mathcal{S} \to \mathsf{BoolVar} \right) \; \times \\
    &\left(\mathsf{Txn}_\mathcal{S} \times \mathsf{Txn}_\mathcal{S}
    \to \mathsf{BoolVar} \right)                  \; \times \\
    &\left(\mathsf{Txn}_\mathcal{S} \times \mathsf{Txn}_\mathcal{S}
    \to \mathsf{BoolVar} \right)
  \end{aligned}
\]
Where \textsf{Txn}, \textsf{Obj}, and \textsf{Val} denote the
unbounded sorts. Given an object $B \in \mathsf{Vars}$,
$B.\mathsf{Txn}$, $B.\mathsf{Obj}$, and $B.\mathsf{Val}$ are used to
access the first three components of $B$, which correspond to the
finite sorts $\mathsf{Txn}_\mathcal{S}$, $\mathsf{Obj}_\mathcal{S}$,
and $\mathsf{Val}_\mathcal{S}$. We refer to the last four components
of $B$ as $B.\mathsf{writes}$, $B.\mathsf{reads}$, $B.\mathsf{so}$,
and $B.\mathsf{co}$, respectively. Each of these is a function for
retrieving individual propositional variables. As an example, the
variable $B.\mathsf{writes}(t_1, x_0, n_2)$ is true in an execution
where transaction $t_1$ writes the value $n_2$ to object $x_0$.

\subsection*{Formula translation}

After declaring the appropriate set of propositional variables for a
given scope, the \textsf{translate} function can be used to convert a
\textsf{FolFormula} to the corresponding propositional logic formula
under that scope.
\[
  \mathsf{translate} : \mathsf{Vars} \times \mathsf{FolFormula} \to
  \mathsf{PropFormula}
\]

Propositional formulas consist of two different categories of syntax:
\[
  \begin{array}{rl}
    p      & \text{propositional variables} \\
    \circ  & \text{binary connectives}
  \end{array}
\]

We define the abstract syntax for propositional logic as follows.
\[
  \begin{array}{ccl}
    \circ   & ::= & \land \mid \lor \mid \Rightarrow \\
    \varphi & ::= & p \mid T \mid F \mid \neg \varphi \mid \varphi \circ \varphi
  \end{array}
\]

We define \textsf{translate} recursively as follows:
\[
  \begin{array}{lcl}
    \mathsf{translate}\left(B, P(c_1, \ldots, c_n)\right)
    & = & B.P(c_1, \ldots, c_n) \\
    \mathsf{translate}\left(B, \neg \varphi\right)                & =
    & \neg \, \mathsf{translate} \left(B, \varphi\right) \\
    \mathsf{translate}\left(B, \varphi \, \circ \, \psi\right)    & =
    & \mathsf{translate} \left( B, \varphi\right) \, \circ \,
    \mathsf{translate} \left(B, \psi\right) \\
    \mathsf{translate}\left(B, \forall x{:}\tau.\, \varphi\right) & =
    & \bigwedge_{d \in B.\tau} \mathsf{translate}\left(B, \varphi[x
    \mapsto d]\right)  \\
    \mathsf{translate}\left(B, \exists x{:}\tau.\, \varphi\right) & =
    & \bigvee_{d \in B.\tau} \mathsf{translate}\left(B, \varphi[x
    \mapsto d]\right)
  \end{array}
\]

\paragraph{Quantifiers} Quantifiers are eliminated and replaced by
finite conjunctions/disjunctions over the respective finite domain.
Here, use the notation $d \in \tau$ to denote iteration through all
the members of sort $\tau$.

\paragraph{Predicates} We assume the formula is closed, that is all
variables  are bound to a quantifier, and are thus replaced by
constants during translation. Consequently, we need only to specify
how to translate predicates applied to a tuple of constants to the
respective boolean variable.

\paragraph{Boolean connectives} Boolean connectives are translated
homomorphically.


\subsection*{The solve method}
The solve method takes in, as its single argument, a formula in
propositional logic. It converts the formula to CNF and passes it as
input to a SAT solver.
\[
  \mathsf{solve} : \mathsf{PropFormula} \to \mathsf{Solution}
\]

\subsection*{Shorthand substitution notation}
Given an instance $I \in \mathsf{Instance}$ produced by a SAT solver
and a set of propositional variables $B \in \textsf{Vars}$ in the
algorithm presentation we use the following special shorthand
notation for variable substitution:
\begin{align*}
  C[\text{co} \mapsto \text{co}_{I}]_{B} & =
  C[\,B.\text{co}(t_1, t_2) \mapsto I(B.\text{co}(t_1, t_2)) \, \mid
  \, t_1, t_2 \in B.\textsf{Txn}\,] \\
  C[H \mapsto H_{I}]_{B}                 & =
  C[\mathit{HistMap}]
\end{align*}
where \textit{HistMap} denotes the following variable mapping:
\begin{gather*}
  \{B.\textsf{writes}(t, x, v) \mapsto I(B.\textsf{writes}(t, x, v))
    \, \mid \, t \in B.\textsf{Txn}, x \in B.\textsf{Obj}, v \in
  B.\textsf{Val}\} \\
  \cup \\
  \{B.\textsf{reads}(t, x, v) \mapsto I(B.\textsf{reads}(t, x, v)) \,
  \mid \, t \in B.\textsf{Txn}, x \in B.\textsf{Obj}, v \in B.\textsf{Val}\} \\
  \cup \\
  \{B.\textsf{so}(t_1, t_2) \mapsto I(B.\textsf{so}(t_1, t_2)) \,
  \mid \, t_1, t_2 \in B.\textsf{Txn}\}
\end{gather*}

In the algorithm explanation, where the set of propositional
variables in question is implicit in the context, we drop the $B$ subscript.

\subsection*{Soundness and completeness}

In this section we prove the soundness and completeness of Isolde's
\textsf{synth} method.

\paragraph{Soundness.}
We formalize the concept of \textit{soundness} as the following property:
\[
  \forall H \ldotp \forall \mathcal{S} \ldotp \forall N \ldotp
  \left(
    \mathsf{synth}(\mathcal{S}, N) = (\mathsf{SAT}, H)
    \Rightarrow
    H \in \mathcal{S} \land H \not\models N
  \right)
\]
Intuitively, Isolde is sound iff every history $H$ it produces is a
valid solution to the synthesis problem, i.e., (1) $H$ belongs to the
given scope, and (2) $H$ is disallowed by $N$.

The first requirement is satisfied by the soundness of our scope
translation mechanisms described above, which should be
self-explanatory. The second requirement holds iff the history
produced by Isolde cannot be used to construct a valid execution
under $N$ with some commit order. This is guaranteed by lines 8 and 9
of the algorithm: the algorithm returns a candidate history $H$ as
its solution iff there is \textit{no} possible commit-order under
which $H$ satisfies $N$ (line 8). Since all valid commit-orders for a
given history are expressible under the SAT scope, this holds without bounds.

\paragraph{Completeness.} We formalize completeness as the following property:
\[
  \left(\exists H \in \mathcal{S} \ldotp H \not\models N\right)
  \,
  \Rightarrow
  \,
  \left(\exists H \ldotp \mathsf{synth}(\mathcal{S}, N) =
  (\mathsf{SAT}, H)\right)
\]
Intuitively, Isolde is complete iff it is guaranteed to find a
satisfying history under scope $\mathcal{S}$ if such history exists.

We can prove this by contradiction. Let us consider there is a
satisfying history $H \in \mathcal{S}$ such that $H \not\models N$
but Isolde produces $\mathsf{UNSAT}$ for that problem. This means
that one of the constraints passed onto the SAT solver during a
candidate-search step (lines 4 and 12 of the algorithm) excludes $H$
from the search space. This means that either (1) the initial formula
$I$ is not satisfied by $H$ or (2) one of the formulas $L_i$ is not
satisfied by $H$. Both of these are impossible: from the definition
of $H \not\models N$, for \textit{every} commit-order $H$ forms an
execution that violates $N$; and (1) $I$ holds for all executions
that violate $N$ (line 3); and (2) each formula $L_i$ constraints the
solution history to form, together with the commit-order taken from
the most recent counterexample, an execution that \textit{violates}
level $N$. Therefore, all the constraints used in candidate-searching
steps hold for any satisfying history.

%% file: sections/appendix/appendix-evaluation.tex
\section{Supplementary Evaluation}\label{app:eval}

\begin{figure*}[ht]
  \centering
  \scalebox{.4}{\input{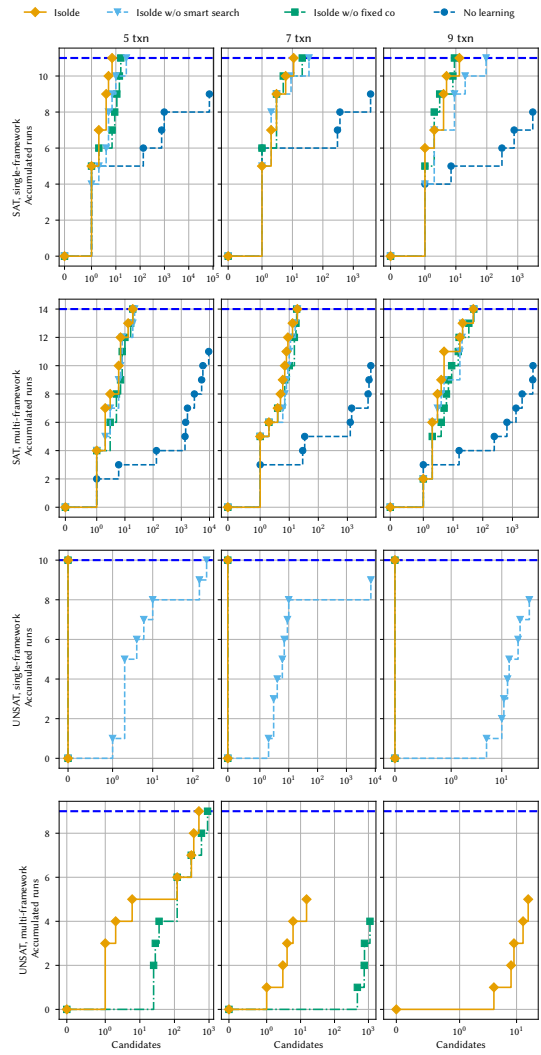}}
  \caption{Cumulative number of benchmark instances solved as a
    function of the total
    number of CEGIS candidates evaluated, broken down by problem type
    (rows) and scope
    (columns). Each curve represents a different implementation; a
    point $(c, k)$ indicates
  that $k$ instances were solved using at most $c$ candidates.}
  \label{fig:cactus-cand}
\end{figure*}

\begin{figure*}[ht]
  \centering
  \scalebox{.4}{\input{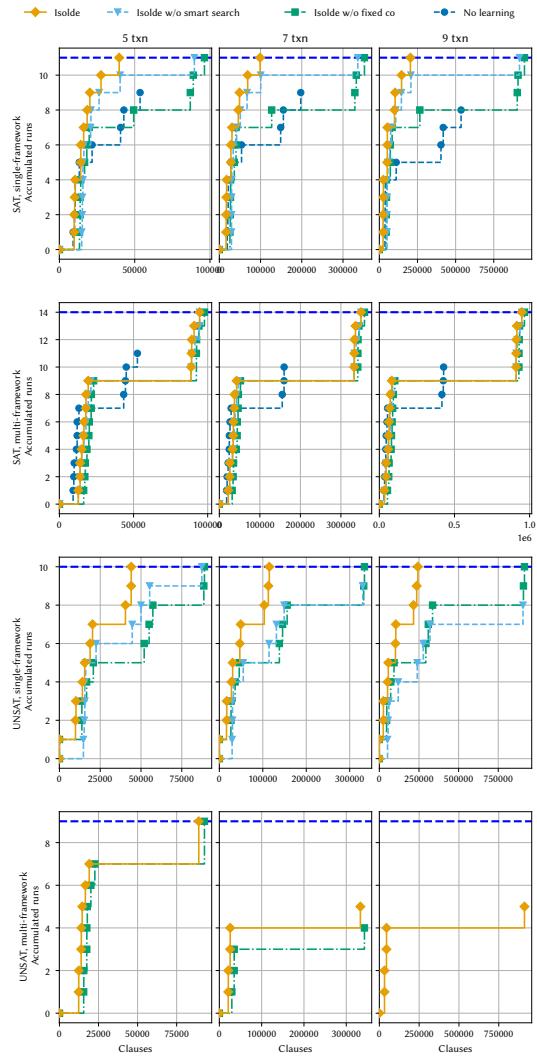}}
  \caption{Cumulative number of benchmark instances solved as a
    function of the number of
    propositional clauses in the initial SAT encoding, broken down by
    problem type (rows) and
    scope (columns). Each curve represents a different
    implementation; a point $(c, k)$
    indicates that $k$ instances were solved with an initial formula of
  at most $c$ clauses.}
  \label{fig:cactus-clauses}
\end{figure*}

\begin{figure*}[ht]
  \centering
  \scalebox{.4}{\input{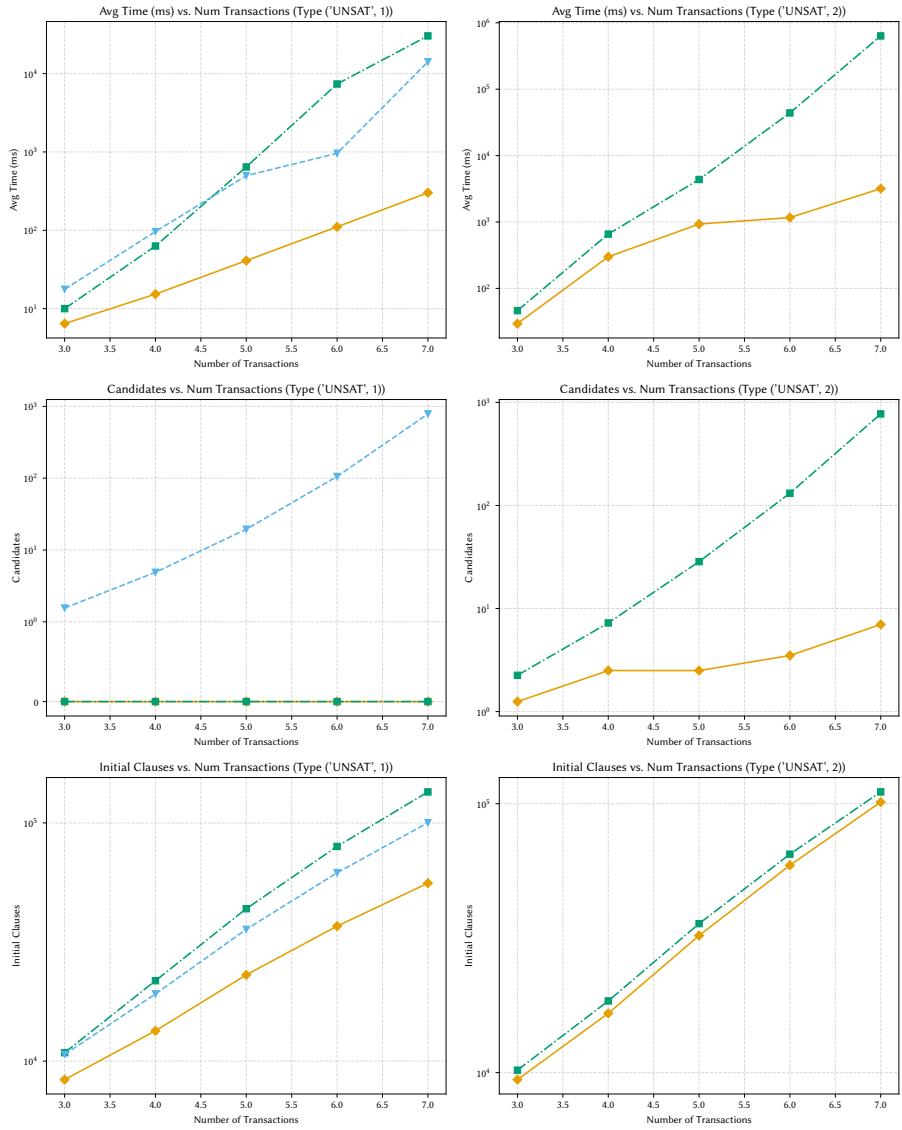}}
  \caption{Effect of scope (number of transactions) on initial clause
    count (top), average
    runtime (middle), and number of CEGIS candidates evaluated (bottom), for
    single-framework (left) and multi-framework (right) UNSAT
    problems. Each line represents
  a different implementation of Isolde.}
  \label{fig:compare-metrics-impl}
\end{figure*}

\begin{figure*}[ht]
  \centering
  \scalebox{.4}{\input{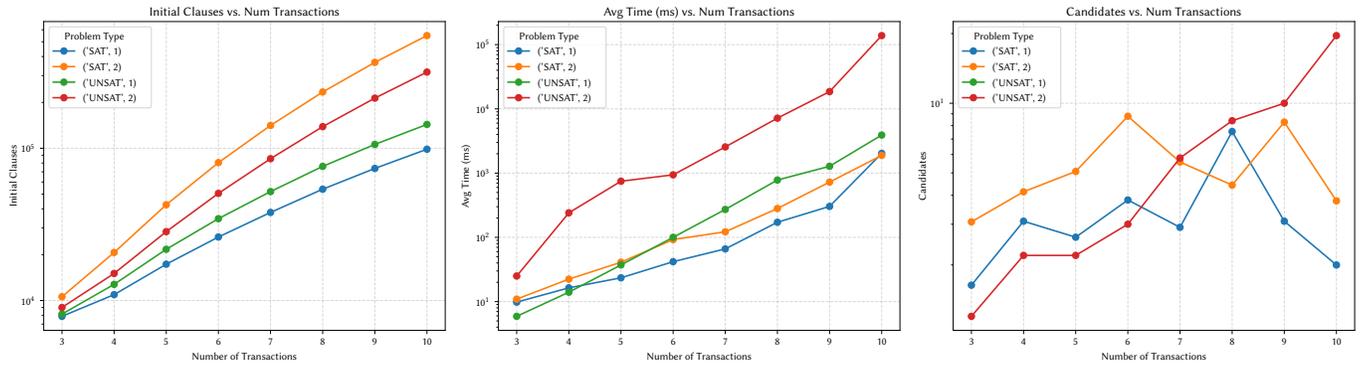}}
  \caption{Effect of scope (number of transactions) on initial clause
    count, average
    runtime, and number of CEGIS candidates (left to right), across
    all four problem types,
  using the full Isolde implementation. Both axes are on a logarithmic scale.}
  \label{fig:compare-metrics-ptypes}
\end{figure*}

This appendix provides additional quantitative detail on the
evaluation presented in
Section~\ref{sec:evaluation}, focusing on two dimensions that
complement the runtime-based
analysis of the main text: (1) the number of CEGIS candidates
evaluated and the size of
the SAT encodings generated, and (2) how these metrics co-vary with
scope and problem type
across implementations. Together,
Figures~\ref{fig:cactus-cand}--\ref{fig:compare-metrics-ptypes}
allow the reader to separate the contribution of better \emph{search
guidance} (fewer
candidates) from the contribution of more \emph{compact encodings}
(fewer clauses), and to
understand how both interact with the transaction scope.

\paragraph{CEGIS candidate counts (Figure~\ref{fig:cactus-cand})}
Figure~\ref{fig:cactus-cand} mirrors the cactus plot structure of
Figure~\ref{fig:cactus-times}
in the main text, but plots cumulative instances solved against the
number of CEGIS
candidates evaluated rather than wall-clock time. A point $(c, k)$ on
a curve means that
$k$ benchmark instances were resolved using at most $c$ candidate histories.

For SAT problems (top two rows), all CEGIS-based variants of Isolde
converge within far
fewer candidates than the no-learning baseline. This is consistent
with our expectations:
the baseline must enumerate candidates without using counterexample
feedback to prune the
search space.

For single-framework UNSAT problems (row 3), the smart search
optimization allows problems
to be resolved in a single solving call, without generating any
candidates at all. Recall
that this optimization merges the two existentially quantified formulas in the
candidate-searching step into a single constraint on one abstract
execution. Because the
formulas characterizing stronger isolation levels imply those
characterizing weaker ones,
the merged candidate-searching formula is immediately unsatisfiable,
and the synthesis
problem is identified as UNSAT without evaluating any candidates. Disabling this
optimization forces Isolde to evaluate several candidates before
reaching the same
conclusion.

For multi-framework UNSAT problems (row 4), disabling the fixed
commit-order optimization
causes some instances to require substantially more candidates than
the full Isolde
configuration. By fixing a total order on transactions upfront, this
optimization eliminates
a significant number of redundant candidates. These results are
consistent with the runtimes
observed in Figure~\ref{fig:cactus-times}. Note that the smart search
optimization does not apply
to multi-framework UNSAT problems and is therefore absent from this row.

\paragraph{SAT formula size (Figure~\ref{fig:cactus-clauses})}
Figure~\ref{fig:cactus-clauses} uses the same cactus structure but
with the $x$-axis
representing the number of propositional clauses in the initial SAT
formula produced by
Kodkod, before any solving begins. This metric reflects the
complexity of the constraint
encoding independently of the solver's runtime behavior.

Despite solving fewer instances, the no-learning baseline does not
produce substantially
larger SAT formulas than the CEGIS-based implementations. This
supports our hypothesis that
Isolde's advantage over the no-learning baseline
(Section~\ref{sec:evaluation}, RQ1) stems
from evaluating fewer candidates, not from more compact encodings.

The two optimizations have a modest effect on formula size for
multi-framework problems,
but yield significantly fewer clauses for single-framework problems.

\paragraph{Scaling of metrics with scope and implementation
(Figure~\ref{fig:compare-metrics-impl})}
The cactus plots in
Figures~\ref{fig:cactus-times}--\ref{fig:cactus-clauses} clearly
visualize performance metrics and failure modes, but make it
difficult to track how these
metrics evolve as the scope grows.
Figure~\ref{fig:compare-metrics-impl} addresses this by
showing how the number of CEGIS iterations and the SAT formula size
scale with the number
of transactions, and how this is ultimately reflected in Isolde's
runtime. Isolde's results
are plotted alongside the ablations to illustrate the impact of each
optimization on
scalability.

These plots focus exclusively on UNSAT problems. SAT problems are
omitted because they are
inherently less challenging --- all implementations solve them
quickly and with little
variation across scope sizes --- making them less informative for
assessing the impact of
the optimizations.

To avoid conflating timed-out instances with those for which concrete
measurements are
available, these plots include only problems for which no
implementation (Isolde or any
ablation) times out. To still retain a representative set, we cap the
scope at seven
transactions, leaving 9 single-framework and 4 multi-framework UNSAT problems.

Both optimizations reduce the size of SAT problems for
single-framework UNSAT instances.
For multi-framework UNSAT instances, the fixed commit-order
optimization has a negligible
effect on formula size. In terms of candidate counts, the smart
search optimization is
most impactful for single-framework UNSAT problems, while the fixed commit-order
optimization is most impactful for multi-framework UNSAT problems.
Together, these
reductions translate to meaningful improvements in Isolde's solving
times: although
growth remains exponential, the optimizations substantially lower the
growth rate.

\paragraph{Cross-problem-type comparison
(Figure~\ref{fig:compare-metrics-ptypes})}
Finally, Figure~\ref{fig:compare-metrics-ptypes} shows how each of
the three metrics
scales with the number of transactions, broken down by problem
category. As before, only
instances for which Isolde produces a definitive answer within the
timeout are included;
multi-framework UNSAT problems that time out are excluded.

The main takeaway is that Isolde's runtime grows exponentially
with the number of
transactions across all problem types, but multi-framework UNSAT
problems grow at a
markedly faster rate than the rest. Formula size exhibits the same trend, with
multi-framework problems requiring more SAT clauses. Candidate counts
grow erratically
with scope for most problem types, but increase exponentially for
multi-framework UNSAT
problems. For single-framework UNSAT problems, Isolde requires no
candidates at all (as
explained above), so this category is omitted from the candidate count plot.

